\documentclass[twocolumn]{aastex61}

\newcommand{\Swift}{\textit{Swift}}

\newcommand{\HST}{\textit{HST}}
\newcommand{\EK}{\ensuremath{E_{\rm K}}}
\newcommand{\EKiso}{\ensuremath{E_{\rm K,iso}}}
\newcommand{\Egamma}{\ensuremath{E_{\gamma}}}
\newcommand{\Egammaiso}{\ensuremath{E_{\gamma,\rm iso}}}
\newcommand{\epse}{\ensuremath{\epsilon_{\rm e}}}
\newcommand{\epsb}{\ensuremath{\epsilon_{B}}}
\newcommand{\dens}{\ensuremath{n_{0}}}
\newcommand{\Astar}{\ensuremath{A_{*}}}

\newcommand{\tjet}{\ensuremath{t_{\rm jet}}}
\newcommand{\thetajet}{\ensuremath{\theta_{\rm jet}}}

\newcommand{\AV}{\ensuremath{A_{\rm V}}}
\newcommand{\AB}{\ensuremath{A_{\rm B}}}

\newcommand{\nua}{\ensuremath{\nu_{\rm a}}}
\newcommand{\nusa}{\ensuremath{\nu_{\rm sa}}}
\newcommand{\nuac}{\ensuremath{\nu_{\rm ac}}}
\newcommand{\numax}{\ensuremath{\nu_{\rm m}}}
\newcommand{\nuc}{\ensuremath{\nu_{\rm c}}}

\newcommand{\nuX}{\ensuremath{\nu_{\rm X}}}

\newcommand{\nuNIR}{\ensuremath{\nu_{\rm NIR}}}

\received{}
\revised{}
\accepted{}
\submitjournal{ApJ}

\shorttitle{The properties of GRB\,120923A at $z\approx7.8$ }
\shortauthors{Tanvir et al.}

\begin{document}

\title{The properties of GRB\,120923A at a spectroscopic redshift of \MakeLowercase{\it{\large z}}~$\approx7.8$}

\correspondingauthor{N.~R.~Tanvir}
\email{nrt3@leicester.ac.uk}

\author[0000-0003-3274-6336]{N.~R.~Tanvir}
\affiliation{Department of Physics and Astronomy, University of Leicester, University Road, Leicester, LE1 7RH, UK}

\author{T.~Laskar}
\affiliation{National Radio Astronomy Observatory, 520 Edgemont Road, Charlottesville, VA 22903, USA}
\affiliation{Department of Astronomy, University of California, 501 Campbell Hall, Berkeley, CA 94720-3411, USA}

\author{A.~J.~Levan}
\affiliation{Department of Physics, University of Warwick, Coventry, CV4 7AL, UK}

\author{D.~A.~Perley}
\affiliation{Astrophysics Research Institute, Liverpool John Moores University, IC2, Liverpool Science Park, 146 Brownlow Hill,  Liverpool, \\ L3 5RF, UK}

\author{J.~Zabl}
\affiliation{Dark Cosmology Centre, Niels Bohr Institute, Copenhagen University, Juliane Maries Vej 30, 2100 Copenhagen \O, Denmark}

\author{J.~P.~U.~Fynbo}
\affiliation{Dark Cosmology Centre, Niels Bohr Institute, Copenhagen University, Juliane Maries Vej 30, 2100 Copenhagen \O, Denmark}

\author{J.~Rhoads}
\affiliation{School of Earth and Space Exploration, Arizona State University, Tempe, AZ 85287, USA}

\author{S.~B.~Cenko}
\affiliation{Astrophysics Science Division, NASA Goddard Space Flight Center, 8800 Greenbelt Road, Greenbelt, MD 20771}
\affiliation{Joint Space-Science Institute, University of Maryland, College Park, MD 20742, USA}

\author{J.~Greiner}
\affiliation{Max-Planck-Institut f\"ur Extraterrestrische Physik, Giessenbachstrasse, D-85748 Garching, Germany}
\affiliation{Excellence Cluster Universe, Technische Universit\"at M\"unchen, Boltzmannstrasse 2, D-85748, Garching, Germany}

\author{K.~Wiersema}
\affiliation{Department of Physics and Astronomy, University of Leicester, University Road, Leicester, LE1 7RH, UK}

\author{J.~Hjorth}
\affiliation{Dark Cosmology Centre, Niels Bohr Institute, Copenhagen University, Juliane Maries Vej 30, 2100 Copenhagen \O, Denmark}

\author{A.~Cucchiara}
\affiliation{University of the Virgin Islands, \#2 John Brewers Bay, 00802 St Thomas, VI, USA}

\author{E.~Berger}
\affiliation{Harvard-Smithsonian Center for Astrophysics, 60 Garden Street, Cambridge, MA 02138, USA}

\author{M.~N.~Bremer}
\affiliation{H.H. Wills Physics Laboratory, University of Bristol, Tyndall Avenue, Bristol, BS8 1TL, UK.}

\author{Z. Cano}
\affiliation{Instituto de Astrof\'{\i}sica de Andaluc\'{\i}a (IAA-CSIC), Glorieta de la Astronom\'{\i}a s/n, E-18008, Granada, Spain}

\author{B.~E.~Cobb}
\affiliation{Department of Physics, The George Washington University, Washington, DC 20052, USA}

\author{S.~Covino}
\affiliation{INAF, Osservatorio Astronomico di Brera, Via E. Bianchi 46, I-23807 Merate (LC), Italy}

\author{V.~D'Elia}
\affiliation{INAF-Osservatorio Astronomico di Roma, Via Frascati 33, I-00040 Monteporzio Catone, Italy}
\affiliation{ASI-Science Data Centre, Via del Politecnico snc, I-00133 Rome, Italy}

\author{W.~Fong
\altaffiliation{Einstein Fellow}}
\affiliation{Steward Observatory, University of Arizona, 933 N. Cherry Avenue, Tucson, AZ 85721 USA}

\author{A.~S.~Fruchter}
\affiliation{Space Telescope Science Institute, 3700 San Martin Drive, Baltimore, MD 21218, USA }

\author{P.~Goldoni}
\affiliation{APC, Astroparticule et Cosmologie, Universite Paris Diderot, CNRS/IN2P3, CEA/Irfu, Observatoire de Paris, \\ Sorbonne Paris CitŽ, 10, Rue Alice Domon et LŽonie Duquet, 75205, Paris Cedex 13, France}

\author{F.~Hammer}
\affiliation{GEPI, Observatoire de Paris, CNRS, 5 Place Jules Janssen, Meudon F-92195, France}

\author{K.~E.~Heintz}
\affiliation{Centre for Astrophysics and Cosmology, Science Institute, University of Iceland, Dunhagi 5, 107 Reykjav'k, Iceland}
\affiliation{Dark Cosmology Centre, Niels Bohr Institute, University of Copenhagen, Juliane Maries Vej 30, 2100 Copenhagen, Denmark}

\author{P.~Jakobsson}
\affiliation{Centre for Astrophysics and Cosmology, Science Institute, University of Iceland, Dunhagi 5, 107, Reykjavik, Iceland}

\author{D.~A.~Kann}
\affiliation{Instituto de Astrof\'{\i}sica de Andaluc\'{\i}a (IAA-CSIC), Glorieta de la Astronom\'{\i}a s/n, E-18008, Granada, Spain}

\author{L.~Kaper}
\affiliation{Anton Pannekoek Institute for Astronomy, University of Amsterdam, Postbus 94249, NL-1090 GE Amsterdam, the Netherlands}

\author{S.~Klose}
\affiliation{Th\"uringer Landessternwarte Tautenburg, Sternwarte 5, 07778 Tautenburg, Germany}

\author{F.~Knust}
\affiliation{Max-Planck-Institut f\"ur Extraterrestrische Physik, Giessenbachstrasse, 85748, Garching, Germany}

\author{T.~Kr\"{u}hler}
\affiliation{Max-Planck-Institut f\"ur Extraterrestrische Physik, Giessenbachstrasse, 85748, Garching, Germany}

\author{D.~Malesani}
\affiliation{Dark Cosmology Centre, Niels Bohr Institute, Copenhagen University, Juliane Maries Vej 30, 2100 Copenhagen \O, Denmark}
\affiliation{DTU Space, National Space Institute, Technical University of Denmark, Elektrovej 327, DK-2800 Lyngby, Denmark}

\author{K.~Misra}
\affiliation{Aryabhatta Research Institute of Observational Sciences (ARIES), Manora Peak, Nainital 263 002, India}

\author{A.~Nicuesa~Guelbenzu}
\affiliation{Th\"uringer Landessternwarte Tautenburg, Sternwarte 5, 07778 Tautenburg, Germany}

\author{G.~Pugliese}
\affiliation{Anton Pannekoek Institute for Astronomy, University of Amsterdam, Postbus 94249, NL-1090 GE Amsterdam, the Netherlands}

\author{R.~S\'{a}nchez-Ram\'{\i}rez}
\affiliation{Instituto de Astrof\'{\i}sica de Andaluc\'{\i}a (IAA-CSIC), Glorieta de la Astronom\'{\i}a s/n, E-18008, Granada, Spain}

\author{S.~Schulze}
\affiliation{Department of Particle Physics and Astrophysics, Weizmann Institute of Science, Rehovot 7610001, Israel }

\author{E.~R.~Stanway}
\affiliation{Department of Physics, Gibbet Hill Road, Coventry, CV4 7AL, UK}

\author{A.~de~Ugarte~Postigo}
\affiliation{Instituto de Astrof\'{\i}sica de Andaluc\'{\i}a (IAA-CSIC), Glorieta de la Astronom\'{\i}a s/n, E-18008, Granada, Spain}
\affiliation{Dark Cosmology Centre, Niels Bohr Institute, Copenhagen University, Juliane Maries Vej 30, 2100 Copenhagen \O, Denmark}


\author{D.~Watson}
\affiliation{Dark Cosmology Centre, Niels Bohr Institute, Copenhagen University, Juliane Maries Vej 30, 2100 Copenhagen \O, Denmark}

\author{R.~A.~M.~J.~Wijers}
\affiliation{Anton Pannekoek Institute for Astronomy, University of Amsterdam, Postbus 94249, NL-1090 GE Amsterdam, the Netherlands}

\author{D. Xu}
\affiliation{CAS Key Laboratory of Space Astronomy and Technology, National Astronomical Observatories, Chinese Academy of Sciences, \\ Beijing 100012,
China}



\begin{abstract}

Gamma-ray bursts (GRBs) are powerful probes of early stars and galaxies,
during and potentially even before the era of reionization. Although the number of GRBs identified at $z\gtrsim6$ 
remains small, they provide a unique window on typical star-forming galaxies at that time, and 
thus are complementary to deep field observations. We report the identification of the optical 
drop-out afterglow of {\em Swift} GRB\,120923A in near-infrared Gemini-North imaging, and derive a redshift of 
$z=7.84_{-0.12}^{+0.06}$ from VLT/X-shooter spectroscopy. 
At this redshift the peak 15-150\,keV luminosity of the burst was $3.2\times10^{52}$\,erg\,s$^{-1}$, and in fact the burst was close to
the {\em Swift}/BAT detection threshold. The X-ray and near-infrared afterglow were also faint, and in this sense
it was a rather typical long-duration GRB in terms of rest-frame luminosity.
We present ground- and space-based follow-up  observations spanning from X-ray to radio,
and find that a standard external shock  model  with a constant-density circumburst environment with density, 
$n\approx4\times10^{-2}$\,cm$^{-3}$ gives a good fit to the data. 
The near-infrared light curve exhibits a sharp break at $t \approx 3.4$\,days in the observer frame,
which if interpreted as being due to a jet corresponds to an opening angle of  $\thetajet\approx5$\,degrees. 
The beaming corrected 
$\gamma$-ray energy is then $\Egamma\approx2\times10^{50}$\,erg, while the beaming-corrected kinetic energy 
is lower, $\EK\approx10^{49}$\,erg, suggesting that GRB\,120923A was a comparatively low kinetic energy event.
We discuss the implications of this event for our understanding of the 
high-redshift population of GRBs  and their identification.

\end{abstract}




\section{Introduction}

The early  galaxies in the universe, born in the first few hundred million years after the Big Bang, 
have been the focus of extensive observational searches in recent years.
The interest is not only in the nature of these primordial collapsed objects, but also in whether
the UV light they emitted was sufficient to have brought about the reionization of the 
intergalactic medium (IGM) \citep[e.g.,][]{Robertson15}.
Since the recent {\em Planck} results suggest a peak of the reionization era at $z\sim$8--9 
\citep{Planck16},
the focus on galaxies in the range  $z=7$--10 has become even more intense.

Directly detecting galaxies at such redshifts, however, is highly challenging due to their 
intrinsic faintness and high luminosity distance; the samples of $z>8$ galaxies in the {\em Hubble} 
Ultra-Deep Field (HUDF) are almost entirely candidates based on photometric redshifts.
Furthermore, although Lyman-$\alpha$ emission has now been detected in one galaxy at  $z=8.7$ \citep{Zitrin15},
the rising neutral hydrogen in the IGM itself increasingly absorbs this 
emission, which likely contributes to the declining Lyman-$\alpha$ detection rates at $z>7$
\citep{Bunker13,Bolton13}.
The highest spectroscopic redshifts for galaxies based on the Lyman-$\alpha$ break are $z\approx7.5$,
in the case of a galaxy benefiting from significant amplification by 
gravitational lensing of a comparatively
bright galaxy \citep{Watson15}, and a surprisingly luminous galaxy recently claimed to be at $z\approx11.1$ \citep{oesch16}.

Long-duration gamma-ray bursts (GRBs) are the most luminous transients known \citep[e.g.,][]{Racusin08}, 
and are unambiguously linked to the core-collapse of massive stars \cite[e.g.,][]{hjorth03,xu13,maselli14}.
Thus, they provide an 
alternative tracer of galaxies in the early universe, and indeed are
currently the only signature we have of individual stars at such distances.
Redshifts can often be measured from afterglow spectroscopy, a method that
benefits from their simple underlying power-law continua against which the
Lyman-$\alpha$ break imprints an unmistakable signature at high-$z$. 
Afterglow spectroscopy also gives information on the metal enrichment in the host galaxies, complementary
to measurements of abundances in ancient stars locally \citep[e.g.,][]{frebel15},
and the neutral fraction in the surrounding IGM \citep{Barkana04,Totani06,Tanvir07,Thoene13,Hartoog15, 
Chornock15,Melandri15}.

The hosts of high redshift GRBs provide a census of primordial star-forming galaxies.
It is likely that a large fraction of all star formation at $z>7$ was occurring in small galaxies too faint to be seen in the HUDF 
\citep[e.g.,][]{Bouwens15}, and similarly challenging even for the {\em James Webb Space Telescope} ({\em JWST}) by $z\sim10$.
Deep searches for high-$z$ GRB hosts can in principle directly constrain this fraction, which is crucial for quantifying the
total contribution of galaxies to the reionization budget
\citep[see e.g.,][for applications of this approach]{Tanvir12,Trenti12,mcguire16}.

Of course, fully exploiting GRBs as high redshift probes also depends on understanding  
the extent of any evolution of the GRB population as a whole over cosmic time, and
whether they preferentially select certain host galaxies or modes of star formation.
Recent studies have found evidence that GRBs follow star formation in a fairly unbiased
way below a threshold of roughly a third Solar to Solar metallicity \citep{Kruehler15,vergani15,perley2016,graham2017}, which bodes well for using
them as tracers of star formation at high redshift.
Other studies have found hints of possible evolution of, for example, shorter rest-frame duration \citep{Littlejohns13}
and narrower jet opening angle \citep{Laskar14}, with increasing redshift, although samples remain 
small and selection effects hard to assess.

To date, the most distant GRBs found have been GRB\,090423, with a spectroscopic
redshift of $z=8.2$ \citep{Tanvir09,Salvaterra09}, and GRB\,090429B, with a photometric redshift
of $z\approx9.4$ \citep{Cucchiara11}, although the latter result could be as low as $z\approx7$
if there is significant dust obscuration in the host.
Here we report the discovery of GRB\,120923A at a  spectroscopic redshift of $z\approx7.8$, corresponding to an
age of the universe of $\approx670$\,Myr, and present
our modelling of its afterglow.

Throughout the paper, we adopt the following values for cosmological parameters: 
$H_0=71$\,km\,s$^{-1}$\,Mpc$^{-1}$, $\Omega_M=0.27$ and $\Omega_{\Lambda}=0.73$. 
All times are in  the observer frame, uncertainties are at the 68\% confidence level ($1\sigma$), unless 
otherwise noted, and magnitudes are in the AB system.

\section{Observations}

\subsection{Swift observations}
\label{sec:Swift}
GRB\,120923A triggered the \Swift\ Burst Alert Telescope \citep[BAT;][]{bbc+05} on 2012 Sep  23
at 05:16:06 UT \citep{Yershov12}. The observed burst duration was $T_{90}=27.2\pm3.0$\,s, with a fluence of 
$(3.2\pm0.8)\times10^{-7}$\,erg\,cm$^{-2}$ \citep[15--150\,keV;][]{Markwardt12}. 
The 1-s peak flux was $F_{15-150\,\rm keV}=4.1\times10^{-8}$\,erg\,cm$^{-2}$\,s$^{-1}$ \citep{lien16},
close to the effective detection threshold of BAT.
The  time-averaged $\gamma$-ray spectrum is well fit by a power law with an exponential cut-off, 
with a photon index of $\Gamma = -0.29\pm1.66$ and peak energy, $E_{\rm peak}=44.4\pm10.6$\,keV 
(both at 90\% confidence). Integrating the BAT (15--150\,keV) spectral model taking $z\approx8$ 
(the evidence for a redshift of this order is presented in Section \ref{sec:z}), and including the effect of statistical uncertainties in all measured 
quantities using a Monte Carlo analysis, we find that the isotropic equivalent $\gamma$-ray energy 
is $\Egammaiso = (4.8^{+6.1}_{-1.6})\times10^{52}$\,erg (1--$10^4$\,keV, rest frame).

The \Swift\ X-ray telescope \citep[XRT;][]{bhn+05} began observing the field at 05:18:26.0 UT, 
140\,s after the BAT trigger, leading to a detection of the X-ray afterglow. The source was 
localized to RA\,$=$ $20^{\rm h}15^{\rm m}10\fs73$, Dec=$+06^{\circ}13{\arcmin}16{\farcs}9$ (J2000), with an 
uncertainty radius of $1\farcs9$ (90\% containment). XRT continued observing the afterglow for 
3.6\,days in photon-counting (PC) mode, with the last detection at $\approx0.6$\,days.

We extracted XRT PC-mode spectra using the on-line tool on the \Swift\ website 
\citep{Evans07,Evans09}\footnote{\url{http://www.swift.ac.uk/xrt\_spectra/00534402}}. 
We used Xspec (v12.8.2) to fit the PC-mode spectrum between $1.7\times10^{-3}$ and 0.67\,days, assuming 
a photoelectrically absorbed power law model (\texttt{tbabs $\times$ pow}) at the redshift of the GRB,
and a Galactic neutral  hydrogen column density of 
$N_{\rm H, MW} = 1.5\times10^{21}\,{\rm cm}^{-2}$ \citep{Willingale13}. 
Our best-fit model has a photon index of 
$\Gamma=1.77\pm0.14$ 
(68\% confidence intervals, estimated 
using Markov Chain Monte Carlo in Xspec; C-stat$ = 84$ for 105 degrees of freedom). 
The data do not  constrain intrinsic absorption within the host galaxy \citep[see also][]{Starling13}. 
In the following analysis, we assume $N_{\rm 
H, int} = 0$ and use the 0.3--10\,keV count rate light curve from the \Swift\ website, together 
with 
$\Gamma = 1.77$, to compute the 1\,keV flux density (Table~\ref{tab_xray}).

\begin{deluxetable}{cccc}
\tabletypesize{\scriptsize} 
\tablecaption{GRB\,120923A: Log of X-ray observations\label{tab_xray}}
\tablewidth{0pt}
\tablehead{
\colhead{$\Delta t_{\rm start}$} & \colhead{$\Delta t_{\rm end}$} & 
\colhead{Flux} & \colhead{Flux density}
\\
\colhead{(hr)} & \colhead{(hr)} & 
\colhead{($10^{-12}$\,erg\,cm$^{-2}$\,s$^{-1}$)} & \colhead{at 1\,keV ($\mu$Jy)} 
}
\startdata
0.042 & 0.079 	& $7.2\pm1.6$ & $0.77\pm0.16$	\\
0.079 & 0.118   & $6.9\pm1.5$ & $0.67\pm0.15$	\\
0.118 & 0.165   & $6.0\pm1.3$ & $0.59\pm0.14$	\\
0.165 & 0.224   & $4.5\pm1.0$ & $0.45\pm0.10$	\\
0.224 & 0.275	& $4.1\pm1.0$ & $0.41\pm0.10$	\\
0.275 & 0.431 	& $2.2\pm0.4$ & $0.22\pm0.04$	\\
9.78 & 90.2 & $0.018\pm0.006$ & $(1.8\pm0.7)\times10^{-3}$ \\
\enddata
\tablecomments{ 
XRT 0.3--10\,keV flux measurements obtained in photon-counting mode.
The start and end time of each observation is relative to the BAT
trigger time of 2012 Sep 23 05:16:06 (UT). The count rate light curve has been converted to a flux 
density at 1\,keV using a photon index of $\Gamma = 1.77$.
}
\end{deluxetable}

\subsection{Ground-based imaging}

\begin{figure*}
\epsscale{1.18}
\plotone{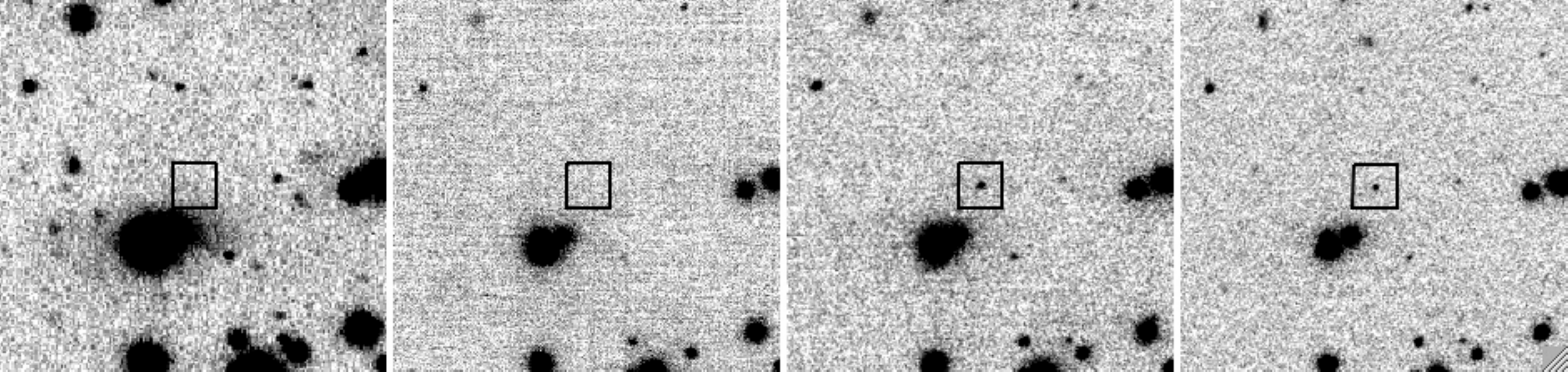}
\caption{Mosaic of GRB\,120923A images obtained with Gemini-North.
Panel 1 is $r+i+z$ from GMOS; panel 2 is $Y$ from NIRI; panel 3 is $J$ from NIRI and panel 4 is 
$H+K$ from NIRI.
The inset box is 3\,arcsec on a side, and the orientation is north up, east left.
\label{fig:mosaic}}
\end{figure*}

We obtained optical and near infra-red (NIR) imaging from Gemini-North using the
Near Infrared Imager and Spectrometer (NIRI) and Gemini Multi-Object Spectrograph (GMOS),
and the United Kingdom Infra-Red Telescope (UKIRT) using the Wide-Field Camera (WFCAM), 
beginning 80\,min after the \Swift\ trigger. 
Conditions in Hawaii were excellent with $\approx0\farcs5$ full-width-half-maximum (FWHM) seeing, and the target was at low airmass for several 
hours. We detected a point source in the $JHK$ bands at RA\,$=20^{\rm h}15^{\rm m}10{\fs}78$, 
Dec\,$=+06^{\circ}13\arcmin16\farcs3$ (J2000), accurate to $\pm0.3\arcsec$ in each dimension,
which is consistent with the X-ray position. The source was absent in the $rizY$ bands 
(Figure~\ref{fig:mosaic}), and its blue colour of $H$-$K\approx0.1$\,mag, together with being a 
$Y$-band drop-out ($Y-J\gtrsim1$\,mag), suggested a very high redshift of $z\gtrsim7$. The NIR 
counterpart faded in subsequent photometry obtained with the Very Large Telescope (VLT)
Infrared Spectrometer and Array Camera (ISAAC), and the 
European Southern Observatory/Max Planck Gesellschaft (ESO/MPG) 2.2m
GRB Optical and Near Infrared Detector \citep[GROND;][]{Greiner08},
in addition to UKIRT and Gemini-North 
over the next several nights, confirming it to be the GRB afterglow.

\begin{deluxetable*}{cclcrcc}
\tabletypesize{\scriptsize}
\tablecaption{GRB\,120923A: Log of optical and NIR imaging observations and  afterglow 
photometry\label{photab}}
\tablewidth{0pt}
\tablehead{
\colhead{$\Delta t_{\rm start}$(hr)} & \colhead{$\Delta t_{\rm end}$ (hr)} & 
\colhead{Telescope/Camera} & \colhead{Filter} & \colhead{Exp. (min)}
& \colhead{Measured flux ($\mu$Jy)} &
\colhead{AB$_0$} 
}
\startdata
1.372 & 1.701 & Gemini-N/NIRI & $J$  & 17  &  $2.74\pm0.14$ &  $22.69\pm0.05$    \\
1.898 & 2.050 & Gemini-N/NIRI & $Y$ & 8  & $0.21\pm0.42$ &  $>23.69$   \\
3.241 & 3.396 & Gemini-N/NIRI & $H$ & 8  &  $2.75\pm0.23$ &  $22.73\pm0.09$  \\
3.442 & 3.597 & Gemini-N/NIRI & $K$ & 8  &  $2.80\pm0.28$  &  $22.73\pm0.10$  \\
3.705 & 4.199 & Gemini-N/NIRI & $Y$ &  26 &  $0.17\pm0.23$ & $>24.26$  \\
4.311 & 4.464 & Gemini-N/NIRI & $J$  & 8  &  $2.09\pm0.33$ & $22.99\pm0.16$  \\
23.79 & 24.14 & Gemini-N/NIRI & $J$  & 17 &  $1.27\pm0.39$ &  $23.52\pm0.29$   \\
2.164 & 2.461 & Gemini-N/GMOS & $r$ & 15  & $0.014\pm0.028$ &  $>26.43$   \\
2.473 & 2.769 & Gemini-N/GMOS & $i$ & 15 &  $0.032\pm0.031$ &   $>26.20$  \\
2.781 & 3.078 & Gemini-N/GMOS & $z$ & 15   &  $-0.029\pm0.150$~\,\,\, & $>25.05$    \\
1.461 & 1.958 & UKIRT/WFCAM & $K$ & 24   &   $4.81\pm0.88$ &  $22.15\pm0.18$   \\
2.036 & 2.538 & UKIRT/WFCAM & $H$ & 24   &   $3.79\pm0.51$ &  $22.38\pm0.14$   \\
2.565 & 3.062 & UKIRT/WFCAM & $J$ & 24   &   $2.38\pm0.62$ &  $22.85\pm0.25$  \\
4.331 & 4.831 & UKIRT/WFCAM & $K$ & 24   &   $3.08\pm1.06$ &  $22.63\pm0.32$  \\
4.852 & 5.225 & UKIRT/WFCAM & $J$ & 18   &   $2.40\pm0.83$ &  $22.84\pm0.32$ \\
23.66 & 24.15 & UKIRT/WFCAM & $H$ & 24   &  $1.16\pm1.10$ &  $>22.51$  \\
18.52 & 18.87 & VLT/ISAAC & $K_{\rm s}$ & 15 & $3.55\pm1.01$ & $22.48\pm0.27$  \\
18.91 & 19.26 & VLT/ISAAC & $H$ & 15 & $2.92\pm0.88$ & $22.66\pm0.29$  \\
19.31 & 19.66 & VLT/ISAAC & $J$ & 15 & $1.57\pm0.56$ & $23.30\pm0.33$ \\
67.82 & 69.25 & VLT/ISAAC & $J$ & 60 & $0.71\pm0.26$ & $24.16\pm0.34$ \\
18.58 & 20.35 & VLT/FORS2 & $z$ & 80 & $0.020\pm0.088$ & $>25.53$  \\
102.2 & 103.9 & \HST/WFC3-IR & F140W & 10 & $0.21\pm0.03$ & $25.49\pm0.12$  \\
120.0 & 127.8 & \HST/WFC3-IR & F140W & 25 & $0.13\pm0.02$ & $26.02\pm0.14$  \\
156.5 & 159.8 & \HST/WFC3-IR & F140W & 15 & $0.11\pm0.02$ & $26.20\pm0.19$  \\
172.5 & 183.7 & \HST/WFC3-IR & F140W & 10 & $0.072\pm0.015$ & $26.66\pm0.21$ \\
477.3 & 479.1 & \HST/WFC3-IR & F140W & 43.5 & $0.014\pm0.010$ & $>27.46$  \\
\enddata
\tablecomments{
The start and end time of each observation is relative to the BAT trigger time of 2012 Sep 23 
05:16:06 (UT). The fluxes are as measured at the location of the afterglow, whereas the AB 
magnitudes are corrected for Galactic foreground extinction \citep[from][]{Schlafly11}, and in cases
of no significant detection are reported as 2$\sigma$ upper limits.}
\end{deluxetable*}

\begin{figure}
\epsscale{1.65}
\plotone{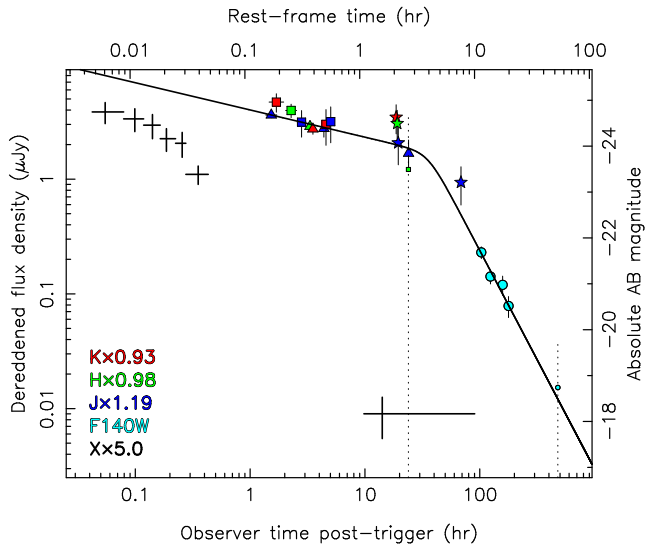}
\caption{Infrared and X-ray light curves of GRB\,120923A.
The NIR measurements have been corrected for extinction by Milky Way foreground absorption using 
$\AB=0.40$ \citep{Schlafly11}, and scaled to overlap as indicated in the key. Each telescope is plotted as a different 
symbol (square - UKIRT; triangle - Gemini-North; star - VLT; circle - \HST); small symbols indicate 
no significant detection at  2$\sigma$. Horizontal bars represent the duration of the observation, 
and vertical error bars are the 1$\sigma$ photometric uncertainties, except in the cases of no 
significant detection, where instead the line is dotted and extends to 2$\sigma$. A smoothly broken 
power-law model is plotted through the data: the early decay is very shallow, with 
$\alpha\approx-0.25$, but the break at about 35\,hr leads to a steep $\alpha\approx-2$ decay
(Note, an afterglow model fit to the full multiwavelength data-set is shown in Figure~\ref{fig:multi}).
The alternative axes showing rest-frame properties assuming a redshift of $z=8$. The black 
points are XRT observations modelled at 1\,keV, and the $x$ and $y$ bars representing the durations 
of the bins and the photometric uncertainties, respectively. The decline rate of the X-ray light 
curve before 10\,h is considerably steeper than that of the NIR light curve over the same 
period. 
\label{fig:lc}}
\end{figure}

We performed photometry using \texttt{Gaia}\footnote{\url{http://astro.dur.ac.uk/$\sim$pdraper/gaia/gaia.html}}, 
with the target aperture placed at the location of the 
afterglow as determined from the high-$S/N$ $J$-band image and the aperture size set according to 
the seeing ($\approx1.3\times$FWHM). We tied the calibration of the WFCAM $JHK$-band images to the 2MASS photometric 
system\footnote{\url{http://www.ipac.caltech.edu/2mass/releases/allsky/doc/sec6\_4a.html}}
using many stars on each frame, and tied the smaller field NIRI and ISAAC images to this using a secondary 
sequence of fainter stars close to the burst location. 
We obtained optical $riz$-band calibration using the wide-field GROND observations.
The $Y$-band calibration was achieved by interpolating the sequence star magnitudes between 
$z$ and $J$ according to $Y = J + 0.534 (z - J)-0.058 $ (derived from GROND observations of photometric standard stars). 
Uncertainties introduced by these calibration steps are included in the error budget, but
are small compared to the random errors on the afterglow photometry.
We summarise our NIR and optical 
observations and photometry in Table~\ref{photab} (note, the GROND limits at the afterglow location are not reported since 
they are shallower than the corresponding VLT observations obtained at almost the same time), 
and present the resulting light curves in 
Figure~\ref{fig:lc}. Constraints on the photometric redshift are outlined in Section~\ref{sec:photoz}.

\subsection{Ground-based spectroscopy}
\label{sect_xs}

Our first spectrum of the afterglow was obtained with the VLT/X-shooter spectrograph \citep{Vernet11}
together with the $K$-band blocking filter, beginning 0.78\,days post-burst 
(Table~\ref{spectab}). The slit width was fixed at $0\farcs9$, which is reasonably matched to the $\approx1\farcs0-1\farcs1$ seeing.
The target was acquired by offsetting from a nearby bright star.

We nodded the target between two positions (A and B, separated by $5\arcsec$) on the slit and took 
exposures in an `ABBA' sequence, as is usual for X-shooter.  This was repeated at two different position
angles of the slit,
specifically 157.6$^{\circ}$ and -161.8$^{\circ}$ (defined from N through E), which maintained an approximately parallactic position.
The data were reduced using the X-shooter pipeline \citep{Goldoni11}.
The spectra were first rectified and 
re-sampled to produce linear spectra on a uniform $0.6$\,\AA\,pix$^{-1}$ wavelength scale. Preliminary sky subtraction was 
done by differencing neighbouring frames. No continuum trace was visible at this stage.
We refined the sky subtraction by masking out the brightest sky lines, and subtracting any residual 
sky signal channel by channel. Atmospheric throughput variations were calibrated by reference to 
observations of two telluric standard stars (Hip094250, Hip094986) obtained close in time to the science data. Channels 
with the highest telluric absorption ($>50\%$), bad pixels, and other image artefacts were all 
masked out. Finally, we co-added all the frames weighted by their respective signal-to-noise ratios,
and optimally binned the data in wavelength to produce wide, 30\,\AA, channels. This revealed a weak but 
clear trace at the expected position on the slit in the NIR arm (which covers wavelength range $\sim1.02-1.8\,\mu$m). We normalized the absolute flux 
scale of the spectrum to match the $J$-band
photometry at the same epoch. No signal was detected in either the UVB ($\sim0.35-0.56\,\mu$m) or VIS ($\sim0.56-1.02\,\mu$m) arms.
The spectroscopic redshift we deduce is presented in Section~\ref{sec:specz}.

\begin{deluxetable*}{rrrcr}
\tabletypesize{\scriptsize}
\tablecaption{GRB\,120923A: Log of spectroscopic observations  \label{spectab}}
\tablewidth{0pt}
\tablehead{
\colhead{$\Delta t_{\rm start}$(hr)} & \colhead{$\Delta t_{\rm end}$ (hr)} & 
\colhead{Telescope/Camera} & \colhead{Spectral element} & \colhead{Exp. (min)}
}
\startdata
18.69 & 21.69 & VLT/X-shooter & - & 160 \\
102.3 & 104.6 & \HST/WFC3-IR & G141 & 80 \\
158.1 & 160.4 & \HST/WFC3-IR & G141 & 80 \\
172.5 & 173.2 & \HST/WFC3-IR & G141 & 40 \\
183.7 & 184.4 & \HST/WFC3-IR & G141 & 40 \\
\enddata
\tablecomments{
The start and end time of each observation is relative to the BAT trigger time of 2012 Sep 23 
05:16:06 (UT).}
\end{deluxetable*}

\subsection{Hubble Space Telescope observations}

We triggered our cycle 19 {\em Hubble Space Telescope} (\HST) program 
to acquire slit-less grism spectroscopy 
of the afterglow with the Wide-Field Camera 3-IR (WFC3-IR), in addition to further imaging in the NIR using the 
F140W filter (approximately a wide $JH$-band). We photometered the afterglow in the F140W images using a $\approx0\farcs32$
radius aperture, adopting the standard \HST\ zero-point calibration and aperture correction for this filter.
This sequence of observations, beginning at $4.3$\,days post-burst, revealed a marked steepening of the light 
curve compared to the previous $J$-band decline rate of $\alpha_{\rm J}\approx-0.25$ 
($F_{\nu}\propto t^{\alpha}$) between $\sim2$\,hr
and $\sim1$\,days (see Figure~\ref{fig:lc} and Section~\ref{sect_ir}). As a consequence of this unexpectedly rapid fading 
of the afterglow, combined with challenges due to overlapping traces from faint sources in the 
crowded field, no usable grism spectrum could be extracted. We report our F140W photometry in Table~\ref{photab} 
and do not consider the grism data further in our analysis.

\begin{figure*}
\begin{center}
\begin{tabular}{ccc}
\includegraphics[width=0.2\textwidth]{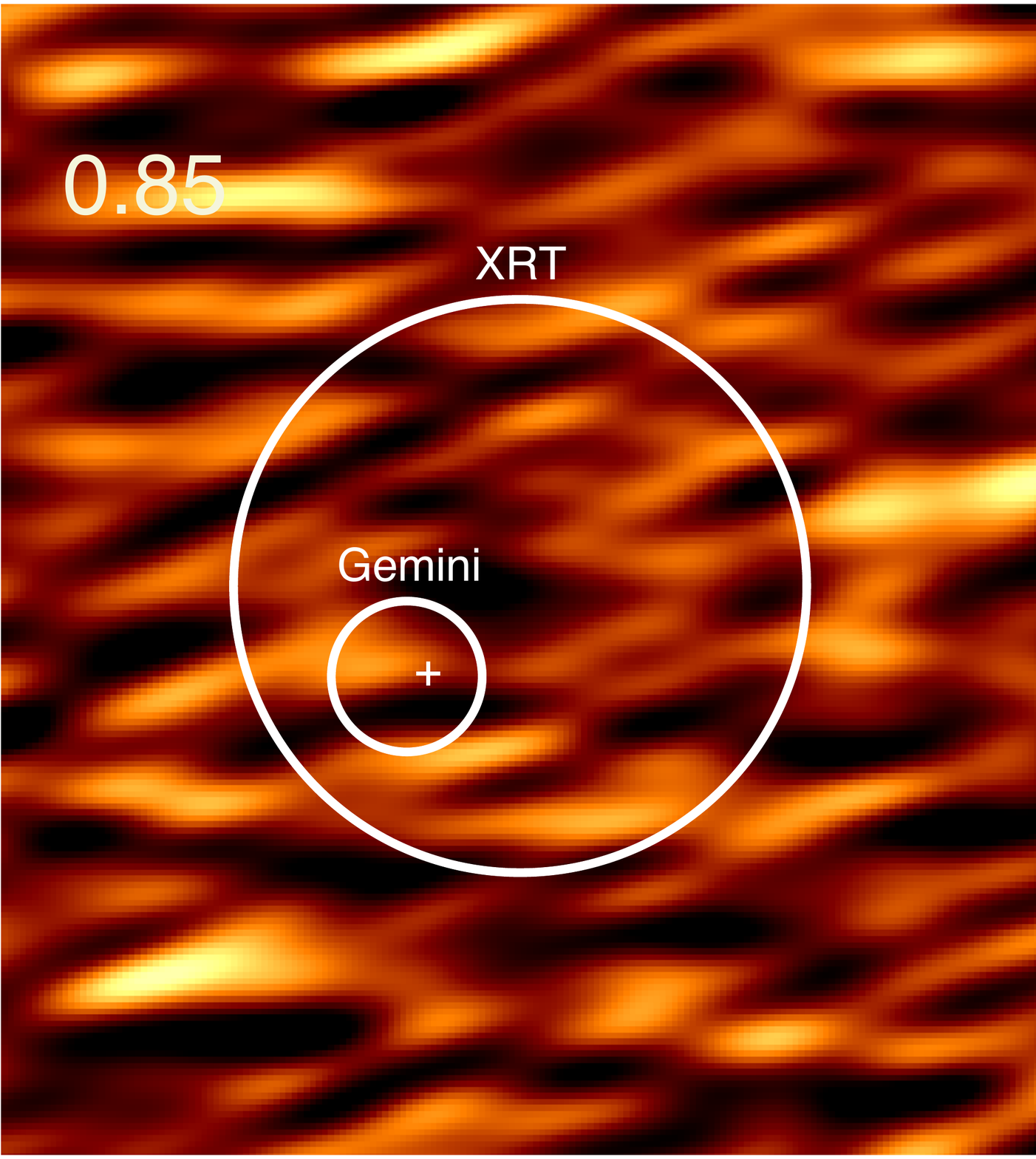} &
\includegraphics[width=0.2\textwidth]{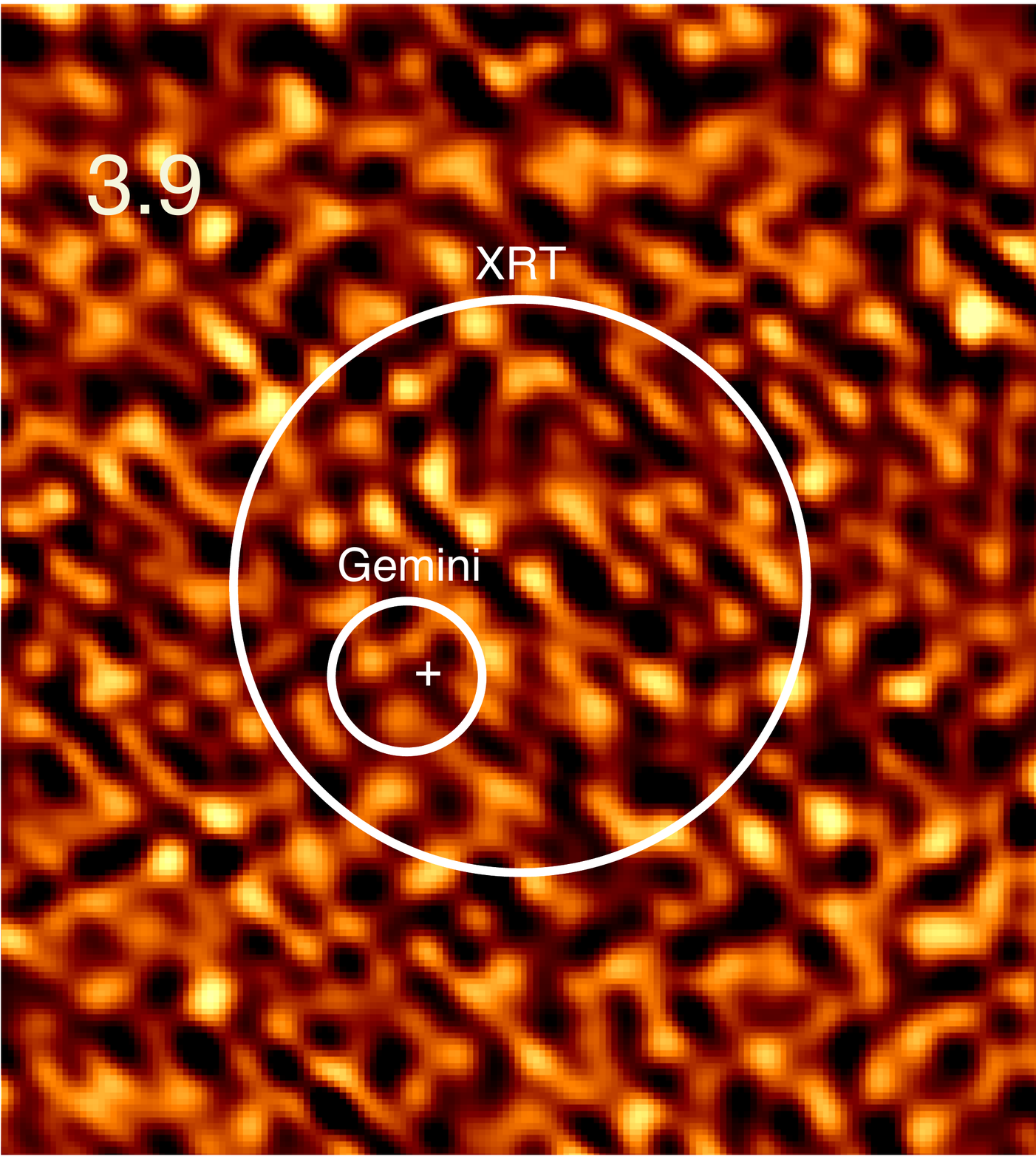} &
\includegraphics[width=0.2\textwidth]{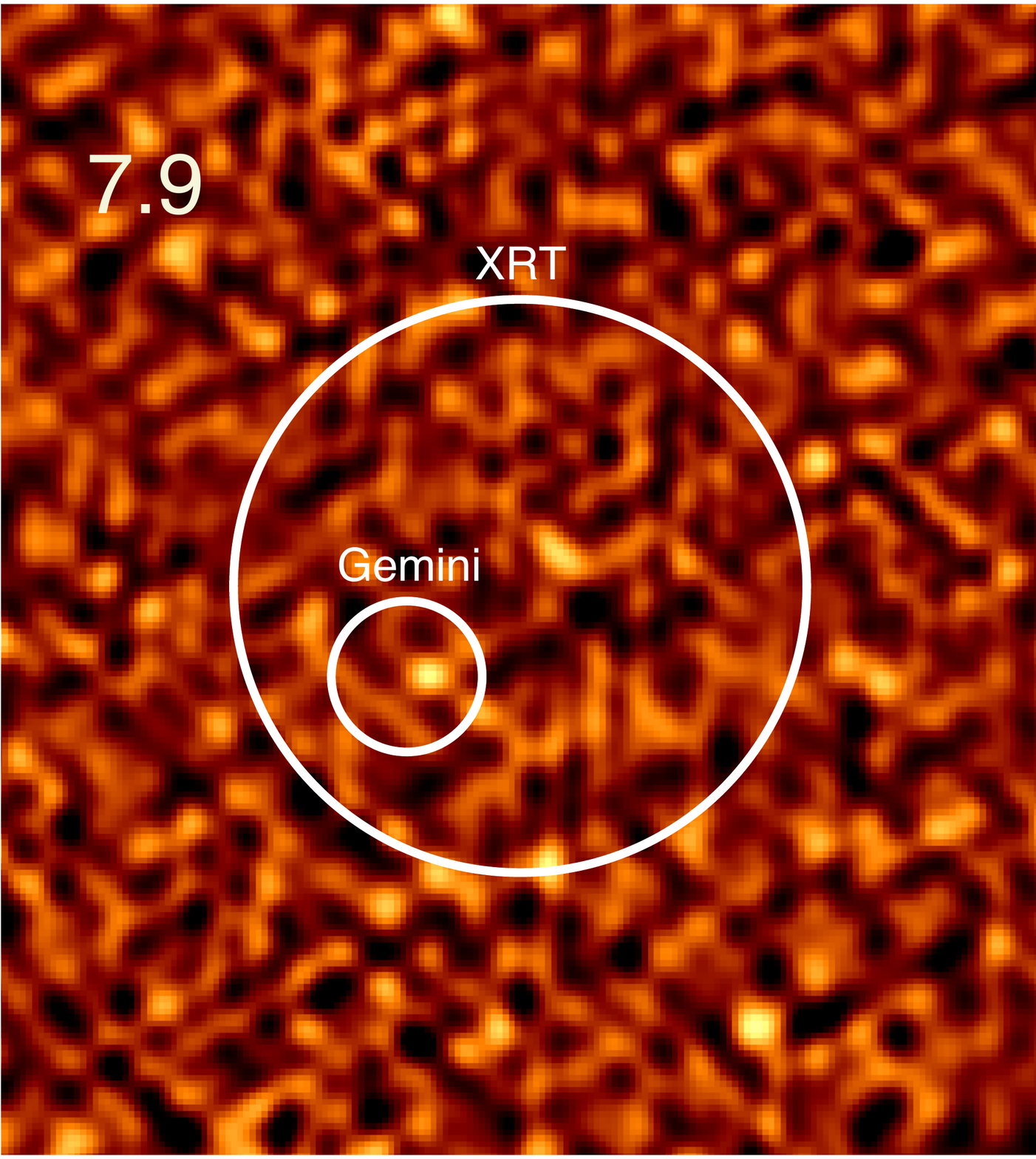} \\ 
 \includegraphics[width=0.2\textwidth]{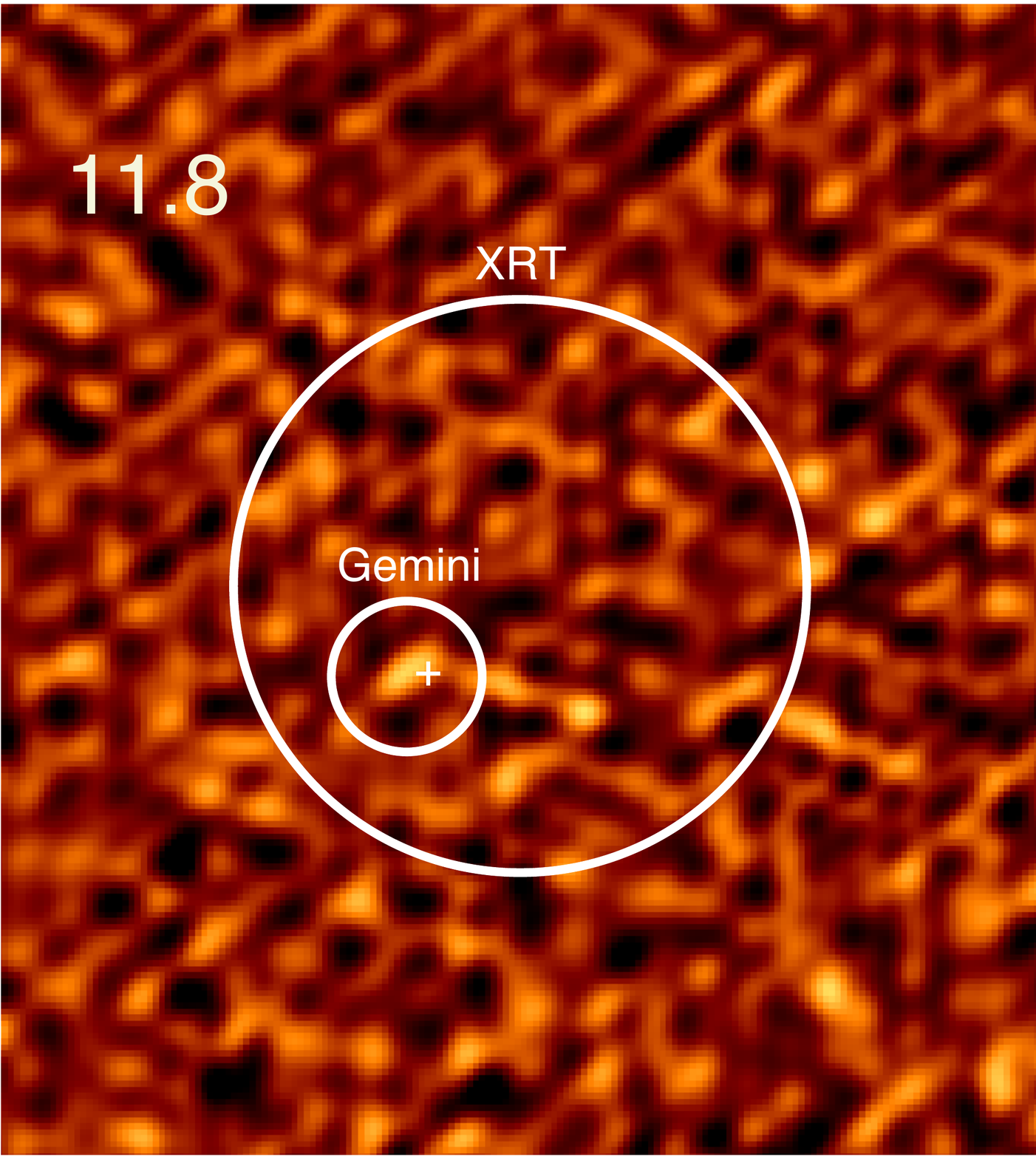} &
 \includegraphics[width=0.2\textwidth]{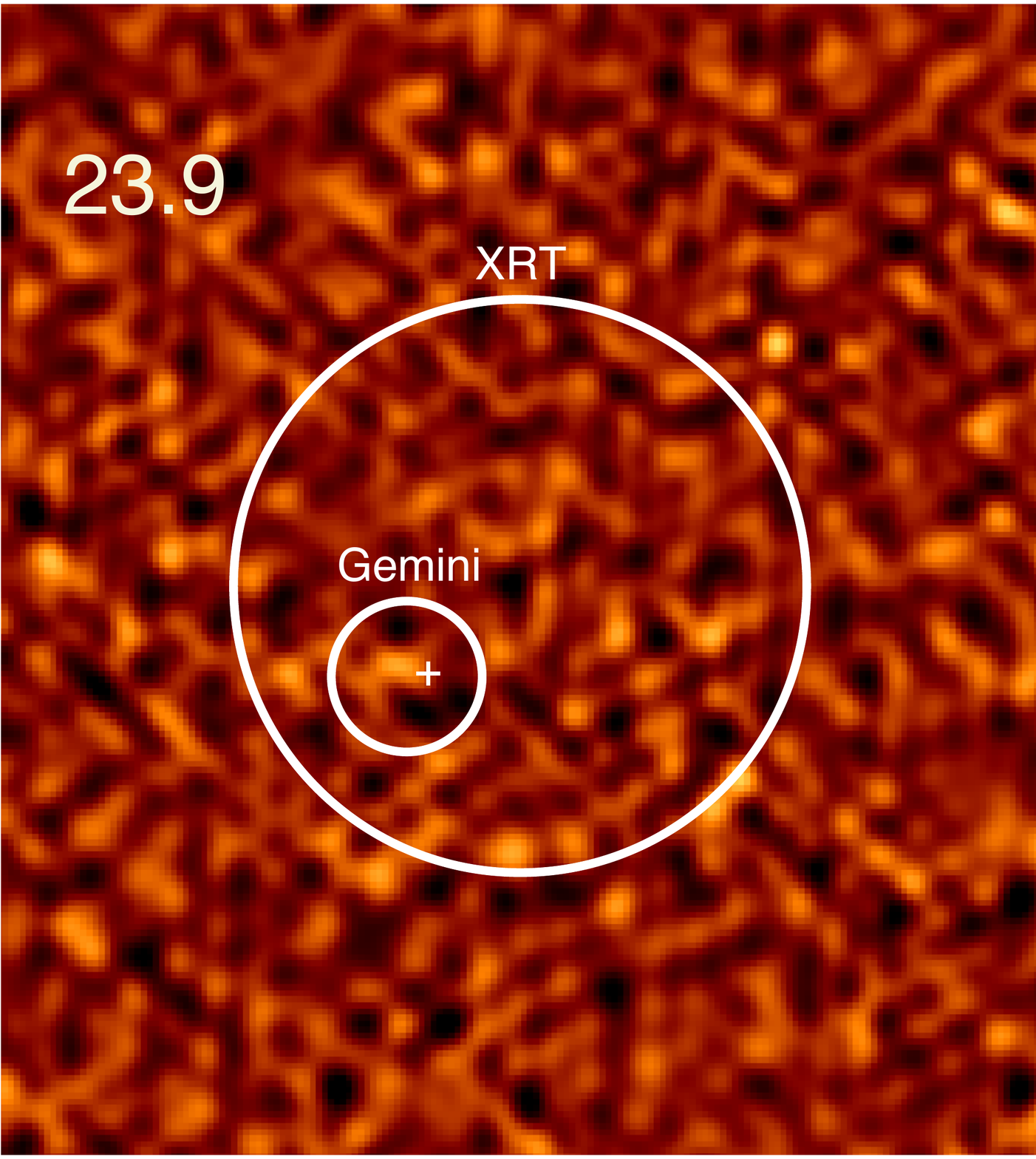} &
 \includegraphics[width=0.2\textwidth]{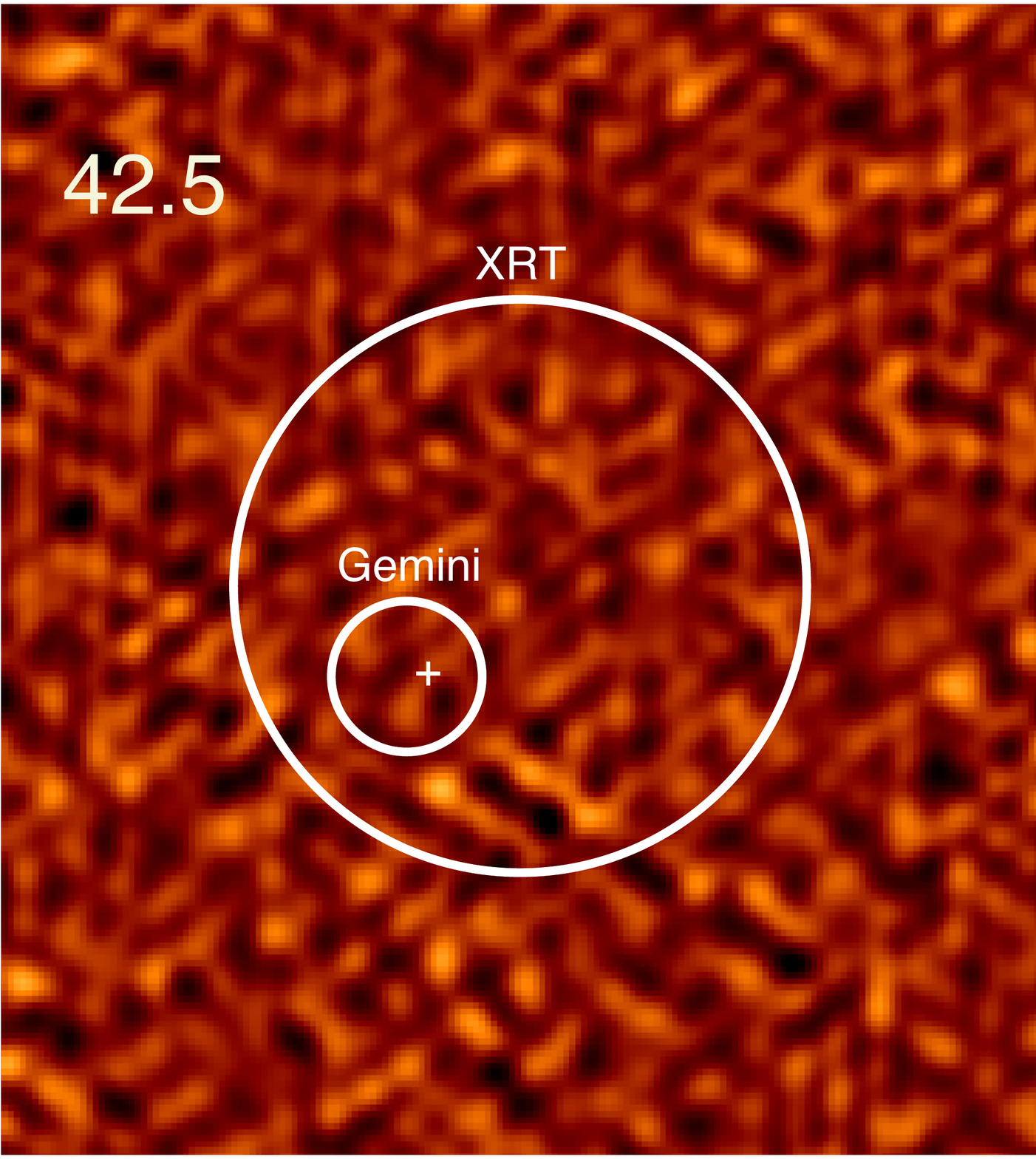} \vspace{-5mm}

\end{tabular}
\end{center}
\caption{6.05\,GHz C band VLA images of the field of GRB\,120923A, together with the XRT and Gemini-North 
error circles. The cross marks the location of the $\approx3\sigma$ possible source in the observation at 
7.9\,days. The position of the peak at 11.8\,d is offset from that at 7.9\,days, suggesting that this is 
not a real radio source. The last image is the stack of observations taken at 40.8\,days and 43.8\,days. We conclude that no 
radio afterglow was detected for this GRB.}
\label{fig:radio}
\end{figure*}

\subsection{Radio observations}

We observed GRB\,120923A with the Combined Array for Research in Millimeter Astronomy (CARMA) 
beginning on 2012 Sep 23.99 UT (0.77 days after the burst) at a mean frequency of 85 GHz. We found no 
significant millimetre emission at the position of the NIR counterpart or within the 
enhanced {\Swift}/XRT error circle to a 3$\sigma$ limit of 0.39\,mJy.

We observed the afterglow in the C (4--7\,GHz; mean frequency 6.05\,GHz) and 
K (18--25\,GHz; mean frequency of 21.8\,GHz) radio bands using the Karl G. Jansky Very 
Large Array (VLA) 
starting 0.82\,days after the burst. 
Depending  on the start time of the observations, we used either 3C286 or 3C48 as the flux and bandpass 
calibrator. We used J1950+0807 as gain calibrator and carried out data reduction using the Common 
Astronomy Software Applications package (\texttt{CASA}\footnote{\url{https://casa.nrao.edu/ }}).

A possible weak source was seen at 7.9\,days in the C band with flux density $25\pm8$\,$\mu$Jy.
We followed up this putative radio afterglow through 
a VLA Director's Discretionary Time proposal (12B-387, PI: Zauderer) over a period of 44\,days (Figure~\ref{fig:radio}). 
The C band observations at 11.8\,days also show a marginally significant peak in the flux density map, but the position
is offset from that of the previous epoch by $\approx2\sigma$. No significant source is detected in 
the subsequent C band epochs nor in the K band data within the Gemini-North error circle. 
Detailed examination and stacking analyses of the images suggests 
that the two possible detections  in the C band are likely due to noise. We therefore consider the VLA 
observations to yield a non-detection of the radio afterglow, and report the upper limits and 
formal  photometric point source fits derived from the maps and stacks in Table \ref{tab:data:radio}.

\begin{deluxetable*}{lcccccc}
\tabletypesize{\footnotesize}
\tablecolumns{9}
\tablewidth{0pt}
\tablecaption{Millimeter and Radio Observations of GRB\,120923A\label{tab:data:radio}}
\tablehead{
  \colhead{Epoch}   &
  \colhead{$t-t_0$} &  
  \colhead{Observatory} &
  \colhead{Band} &
  \colhead{Frequency} &
  \colhead{Integration Time} &
  \colhead{$3\sigma$ Upper Limit}\\
                   &
  \colhead{(days)} & 
                   &
                   &
  \colhead{(GHz)}  &
  \colhead{(min)}  &
  \colhead{($\mu$Jy)}
  }
\startdata
1 & 0.77 & CARMA &   3\,mm     & 85.0 &\ldots& $<390$\\
1 & 0.824 & VLA &  \textit{K} & 21.8 & 17.7 & $<69.1$\\
1 & 0.853 & VLA &  \textit{C} & 6.05 & 19.9 & $<29.7$\\
2 & 3.90  & VLA &  \textit{K} & 21.8 & 18.1 & $<64.4$\\
2 & 3.91  & VLA &  \textit{C} & 6.05 & 17.4 & $<31.0$\\
3 & 7.91  & VLA &  \textit{C} & 6.05 & 27.8 & $<22.5$\\
4 & 11.8  & VLA &  \textit{C} & 6.05 & 35.5 & $<20.1$\\
5 & 23.9  & VLA &  \textit{C} & 6.05 & 37.9 & $<17.1$\\
6 & 40.8  & VLA &  \textit{C} & 6.05 & 36.3 & $<18.4$\\
7 & 43.8  & VLA &  \textit{C} & 6.05 & 49.5 & $<15.8$\\
3\&4$^*$ & 10.1$^\dag$  & VLA &  \textit{C} & 6.05 & 63.4 & $<15.5$\\
6\&7$^*$ & 42.5$^\dag$  & VLA &  \textit{C} & 6.05 & 85.8 & $<14.8$\\
\enddata
\tablecomments{$^*$ Stacks. $^\dag$The reported value of $t-t_0$ for stacks is weighted by the 
integration time of the individual observations used in the stack.}
\end{deluxetable*}

\section{Redshift determination}
\label{sec:z}

\subsection{Photometric redshift constraints}
\label{sec:photoz}
We first investigate the redshift constraints from the optical-NIR spectral energy distribution (SED) of the 
afterglow, using techniques similar to those described in \cite{Laskar14}. For uniformity, we 
selected observations from a single telescope for this analysis, specifically the Gemini-North/GMOS $riz$ 
measurements and the Gemini-North/NIRI $YJHK$ measurements obtained within 5\,hr post-burst. 
We corrected these data for Galactic extinction, using $A_{\rm V,gal}=0.4$\,mag 
\citep{Schlafly11}, and the Milky Way extinction model of \cite{pei92}. The photometry was interpolated  to 
a common time corresponding to the NIRI $H$-band observation, using a power law fit to the $J$-band 
light curve between $0.06$\,days and 1.0\,days, the latter yielding $\alpha_{\rm J} = 
-0.252\pm0.022$. We added the interpolation uncertainty in quadrature with the photometric uncertainty 
to determine the total uncertainty at each point on the SED.

We assumed the intrinsic spectrum of the afterglow is a power law, $F_{\nu} \propto \nu^{\beta}$,
and used the sight-line-averaged model for the optical depth of the IGM  from 
\cite{mad95}, accounting for Ly$\alpha$ absorption by neutral hydrogen 
and  photoelectric absorption by intervening systems. We also included Ly$\alpha$ absorption by the host 
galaxy interstellar medium (ISM), for which we assumed a column density of $\log(N_{\rm H}/{\rm cm}^{-2}) = 21.1$, the mean value for 
GRBs at $z\sim2$--3 \citep{fjp+09}, although within the errors the photometric redshift is insensitive to the exact value chosen. 
The free parameters in our model are the redshift of the GRB, 
the extinction along the line of sight within the host galaxy ($A_V$), and the spectral index 
($\beta$) of the afterglow SED. 
The SMC dust extinction law of
 \citet{pei92} was assumed to model the extinction in the host galaxy, \AV \citep[this is
likely to be appropriate for the expected low metallicity host galaxy, and has also often found
to be a good approximation to the extinction laws in the majority of lower redshift GRB hosts e.g.,][]{schady12,perley13}.
We took a flat prior for the redshift and the extinction, and
employed the distribution of extinction-corrected optical-NIR spectral slopes, $\beta_{\rm o}$ from 
\citet{gkk+11} as a prior on $\beta$. 

\begin{figure}
 \includegraphics[width=\columnwidth]{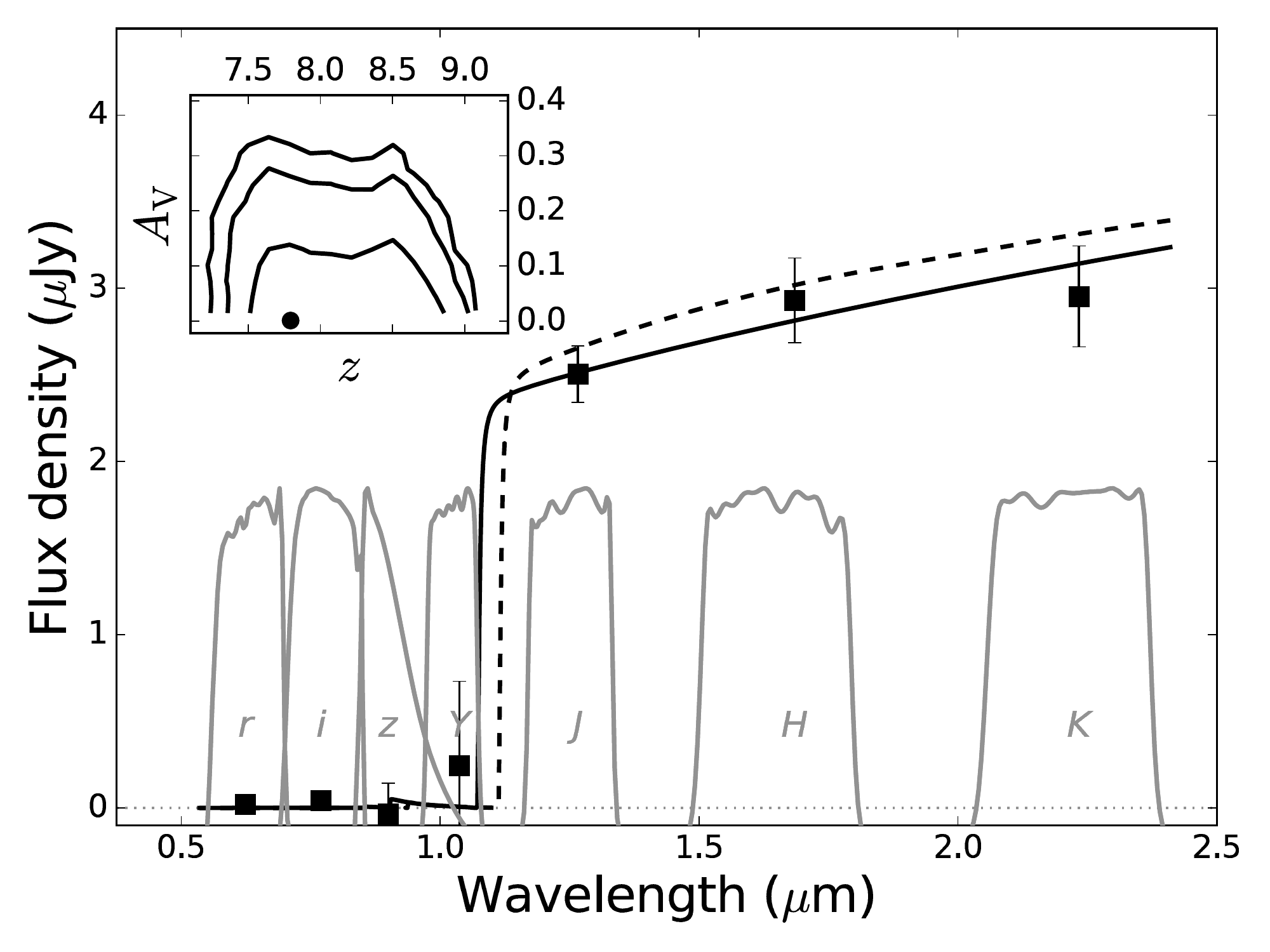}
\caption{The optical-to-NIR spectral energy distribution of GRB\,120923A at 0.14\,days. The 
$JHK$ photometry has been extrapolated from the nearest Gemini-North detections using a fit to the 
$J$-band light curve between 1.4\,h and 24\,h ($\alpha_{\rm J} = -0.252\pm0.022$). The $rizY$ points 
are from forced photometry on Gemini-North images in the same time interval, also interpolated 
using the $J$-band light curve. All photometry (including the $z$-band formal negative flux 
measurement) has been multiplied by a factor $>1$ to correct for Galactic extinction 
\citep[$\AB=0.4$;][]{Schlafly11}. The data points have been placed at the centroid of the filter 
bandpasses for clarity. The lines are models for the afterglow SED, including IGM and ISM 
absorption using the highest-likelihood model (solid) and the median values of the parameter 
distributions (dashed). We show the $1\sigma$, $2\sigma$, and $3\sigma$ correlation contours 
between extinction ($\AV$) and redshift ($z$) in the inset, where the black dot indicates the 
highest-likelihood model.}
\label{fig:sed}
\end{figure}

\begin{figure}
 \includegraphics[width=\columnwidth]{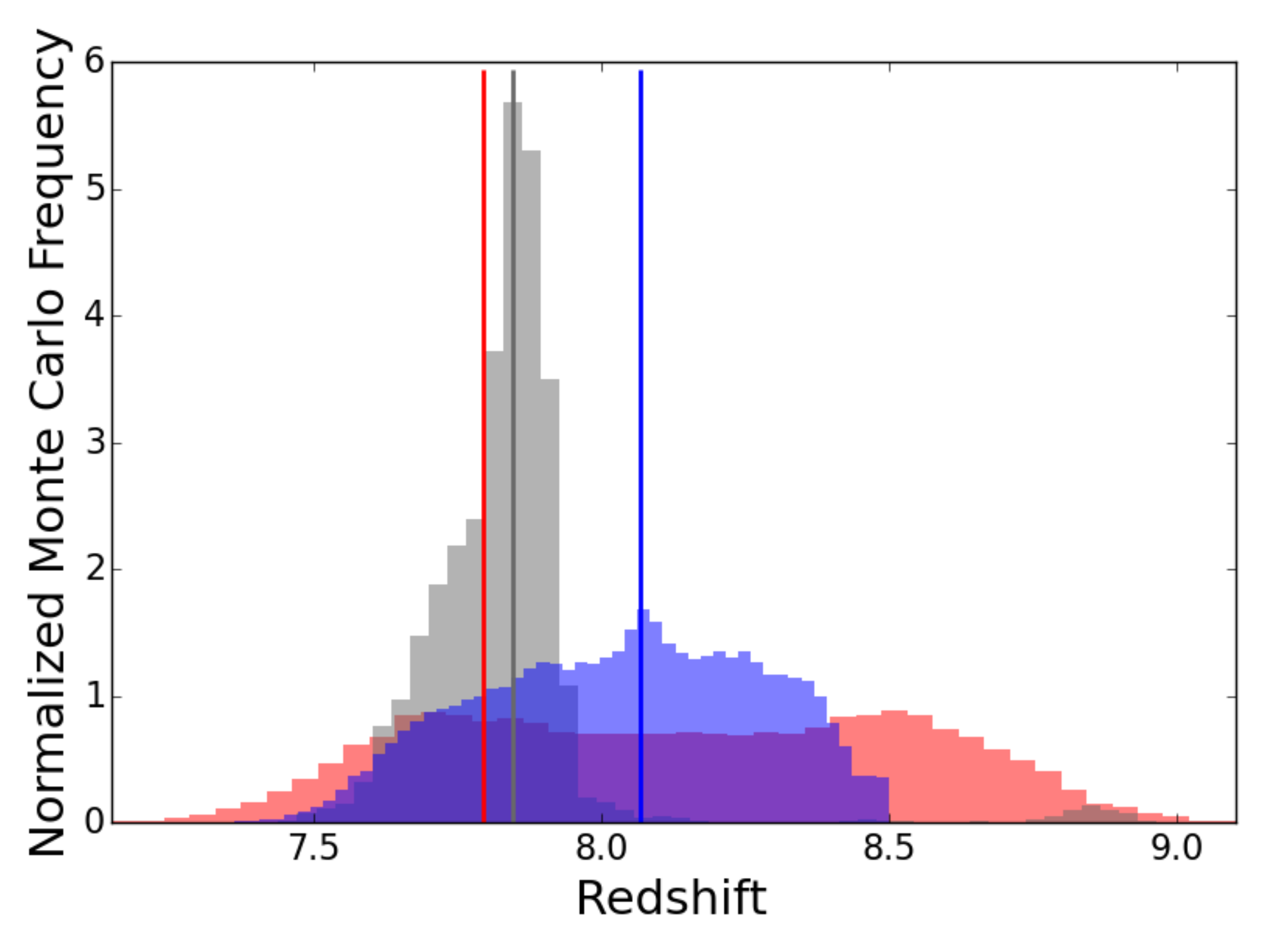}
\caption{Posterior density function for the redshift of GRB\,120923A from fitting the photometric SED
(red, Figure~\ref{fig:sed} and Section~\ref{sec:photoz}), from fitting the NIR spectrum (grey, 
Figure~\ref{fig:xs_1D} and Section~\ref{sec:specz}), and from fitting all the available X-ray to radio afterglow 
data (but not the X-Shooter spectrum) with a physical multi-wavelength model (blue; Section~\ref{sec:multi}). The vertical lines 
indicate the redshifts of the respective best-fit models.}
\label{fig:zhist}
\end{figure}

Fitting was performed using a Markov Chain Monte Carlo (MCMC) algorithm to explore 
the parameter space, integrating the model over the filter bandpasses, and computing the likelihood 
of the model by comparing the resulting fluxes with the observed values using a \texttt{Python} 
implementation of the ensemble MCMC sampler \texttt{emcee} \citep{fhlg13}.
The resulting 68\% confidence intervals about the median values for the fitted parameters are 
$z = 8.1\pm0.4$, $\beta = -0.17^{+0.34}_{-0.25}$, and $\AV = 0.07^{+0.09}_{-0.05}$\,mag,
where the large errors on $\beta$ reflect the limited lever arm obtained from the three $JHK$ detections.
The highest-likelihood model is $z \approx 7.79$, $\beta \approx -0.39$, and $\AV \lesssim 
0.1$\,mag, and this is shown, along with
 a model with the median parameters,
in  Figure~\ref{fig:sed}.
The full posterior density function for the redshift is shown in 
Figure~\ref{fig:zhist}, and allows us to
rule out a redshift of $z\lesssim7.3$ at 99.7\% confidence. 

\subsection{Spectroscopic redshift}
\label{sec:specz}
\begin{figure}
\epsscale{1.19}
\plotone{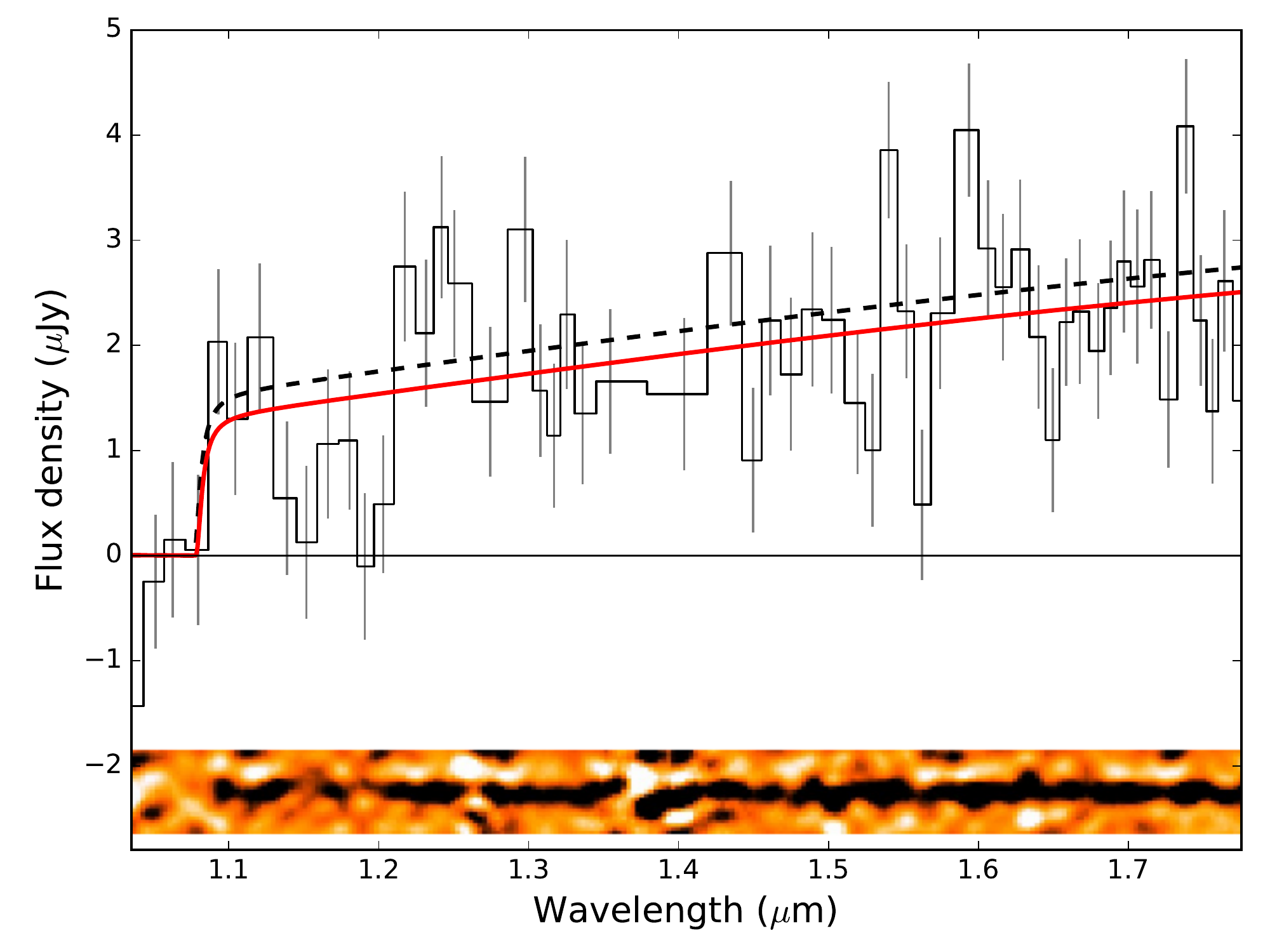}
\caption{VLT/X-shooter spectrum of the GRB\,120923A afterglow. The 1D data have been adaptively 
binned so that the noise in each bin is approximately the same. The 2D spectrum has been re-binned 
and smoothed to enhance the trace of the afterglow. The lines are models for the afterglow 
spectrum, including IGM and ISM absorption using the highest-likelihood model (red, solid) and the 
median values of the parameter distributions (dashed).
\label{fig:xs_1D}}
\end{figure}

The X-shooter spectrum (Figure~\ref{fig:xs_1D}) exhibits significant flux redward of $1.2\,\mu$m (below
$2.5\times10^{14}$\,Hz), with a spectral slope of $\beta=-0.6\pm0.5$, and a steep cut-off blueward  of 
$\approx1.1\,\mu$m. We model the spectrum as a power law with index $\beta$, and interpret the break 
as due to Ly$\alpha$ absorption by neutral hydrogen in the host galaxy followed by a Gunn-Peterson 
trough blueward of the host absorption. We proceed with the remainder of the analysis as for the 
photometric redshift, assuming a flat prior for the redshift  and the extinction, and again using the 
distribution of $\beta_{\rm o}$ from \cite{gkk+11} as a prior on $\beta$. 
We also fix the neutral hydrogen column density of the host galaxy to 
$\log({N_{\rm HI, host}/{\rm cm^{-2}}})=21.1$ (Section \ref{sec:photoz}); a significantly higher column than 
this is unlikely given the evidence for low extinction and the suggestion of a trend toward somewhat lower columns seen in
GRBs at $z\gtrsim6$ \citep{Chornock15}, whilst assuming a lower value for $N_{\rm HI, host}$ does not change the
derived redshift within the errors.

However, instead of integrating over filter bandpasses, we fitted the model directly to the observed X-shooter NIR spectrum. 
We find $z=7.84^{+0.06}_{-0.12}$, $\beta = -0.54\pm0.40$, and 
$\AV=0.17^{+0.09}_{-0.12}$, where the uncertainties reflect $68\%$ credible intervals about the 
median. We plot our best-fit model in Figure~\ref{fig:xs_1D}. The best fit parameters are 
$z\approx7.8$, $\beta\approx-0.54$, and $\AV\approx0.17$, all consistent with the median values, and 
with the photometric redshift of $z=8.1\pm0.4$ (Section \ref{sec:photoz}).

\section{Burst properties and comparison to the long-GRB population}
\subsection{High energy behaviour}



\begin{figure}
\epsscale{1.27}
\plotone{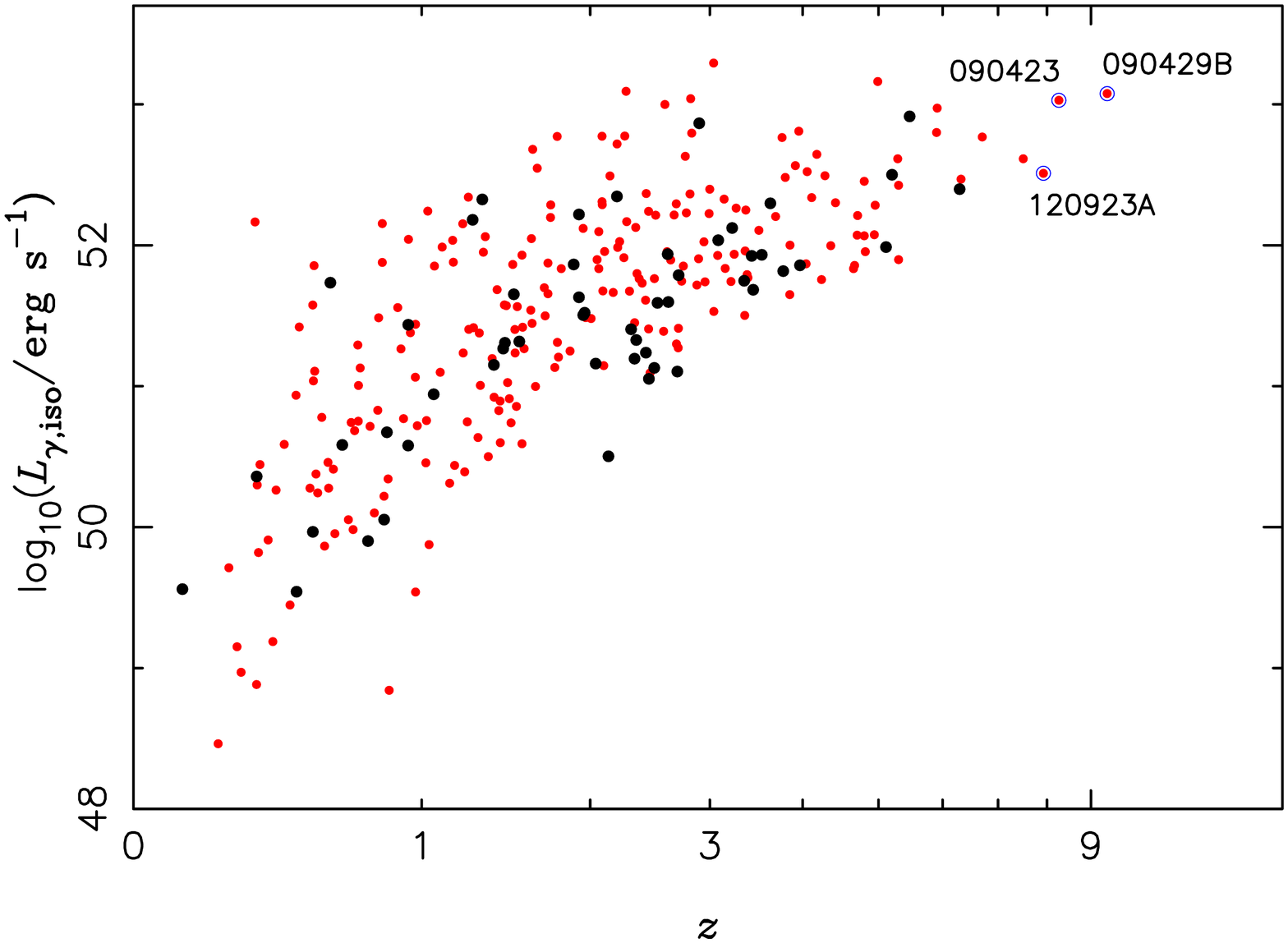}
\caption{
The 1\,s peak  fluxes in the observed 15--150\,keV band \citep{Sakamoto11,lien16}  converted to  rest-frame 
isotropic-equivalent luminosities 
for all \Swift\ bursts with redshifts to March 2015.
The bursts from the
TOUGH sample \citep{Hjorth12,Jakobsson12,Kruehler12} are black points, and the three highest 
redshift bursts are individually labelled.
The effective selection function imposes the lower envelope to this
distribution, although note that in practice the sensitivity of BAT depends on the
instantaneous background count rate,  the location of the burst within its
field of view and the structure of the light curve.
\label{fig:L_vs_z}}
\end{figure}

At $z\approx7.8$, the BAT peak flux corresponds to a luminosity, $L_{\rm iso}\approx3.2\times10^{52}$\,erg\,s$^{-1}$.
In Figure~\ref{fig:L_vs_z} we show the peak luminosity for all the \Swift\ GRBs with measured
redshifts to March 2015. The low energy cut-off imposed by the BAT selection function indicates that only GRBs at 
the bright end of the luminosity function can be detected at $z>6$, despite \Swift\ utilizing a variety 
of algorithms to try to recover even time-dilated bursts \citep[e.g,][]{lien14}.  It is clear that GRB\,120923A was close to
this detection limit, and the intrinsically faintest event found at $z>6.5$ to-date.

\begin{figure}
\epsscale{2.3}
\plotone{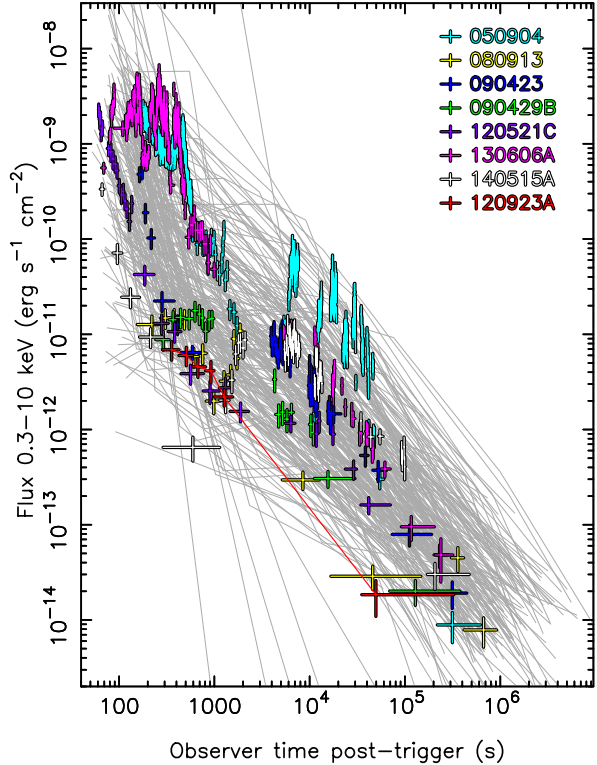}
\caption{Multiply-broken power-law fits to the light curves in the 0.3--10\,keV X-ray band for a large sample of \Swift\ long
duration GRBs \citep{Evans09} in grey, and high-$z$ GRBs (GRB\,130606A at $z=5.9$, GRB\,120521C at $z\approx6.0$,
GRB\,140515A at $z=6.3$, GRB\,050904 at $z=6.3$,
GRB\,080913 at $z=6.7$, GRB\, 090423 at $z=8.2$, and 090429B at $z\approx9.4$) 
in colour.  Data 
for GRB\,120923A are shown in red points, illustrating that it was amongst the faintest afterglows 
seen by \Swift, and factors of several fainter at late times, compared to other $z\gtrsim6$ events.
\label{fig:xrays}}
\end{figure}

\begin{figure}
\epsscale{2.34}
\plotone{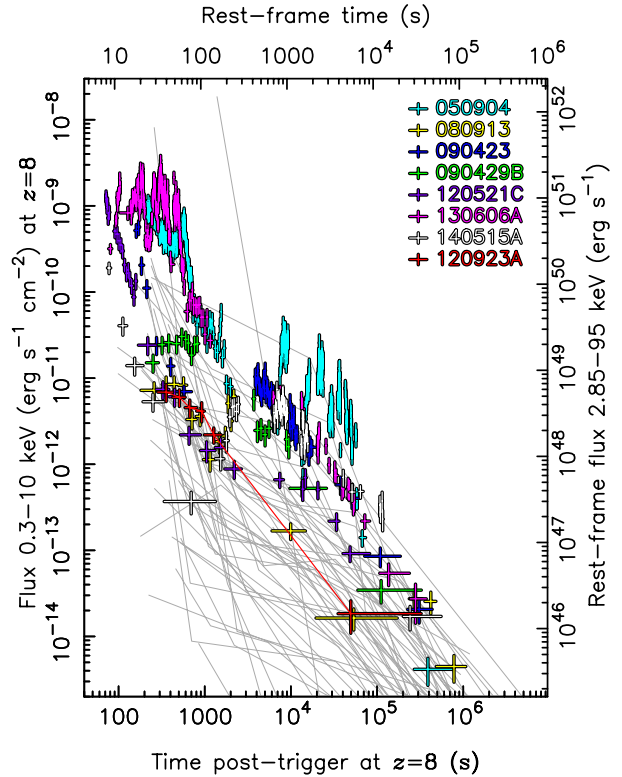}
\caption{Multiply-broken power-law fits to the X-ray light curves of the TOUGH sample of GRBs \citep{Hjorth12}, in grey, together with a selection
of very high-$z$ bursts (see caption to Fig.~\ref{fig:xrays}), redshifted to 
show how they would appear at $z=8$ observed by \Swift. In terms of its intrinsic X-ray light curve 
(see alternative axes), GRB\,120923A is more typical of the TOUGH sample, whilst GRB\,090423, and 
particularly GRB\,050904 would have been amongst the brightest 
in the rest frame.
\label{fig:xrays2}}
\end{figure}

In Figure~\ref{fig:xrays}, we compare the X-ray light curve of GRB\,120923A to those of a large 
sample of \Swift\ bursts \citep{Evans09}. We find that GRB\,120923A was amongst the faintest 
long-duration GRB afterglows seen by XRT.
Another view of these data is shown in Figure~\ref{fig:xrays2}, in which each burst has 
been shifted to show how it would have appeared if it were at $z=8$ (we also show the
corresponding rest-frame axes).  In this case we restrict the low redshift sample to the events 
included in The Optically-Unbiased GRB Host survey  \citep[TOUGH;][]{Hjorth12}. The high redshift 
completeness of that sample minimises any optical selection biases. This shows that GRB\,120923A was 
rather typical in terms of its intrinsic X-ray behaviour.

\subsection{Infrared behaviour}
\label{sect_ir}

\begin{figure}
\epsscale{2.74}
\plotone{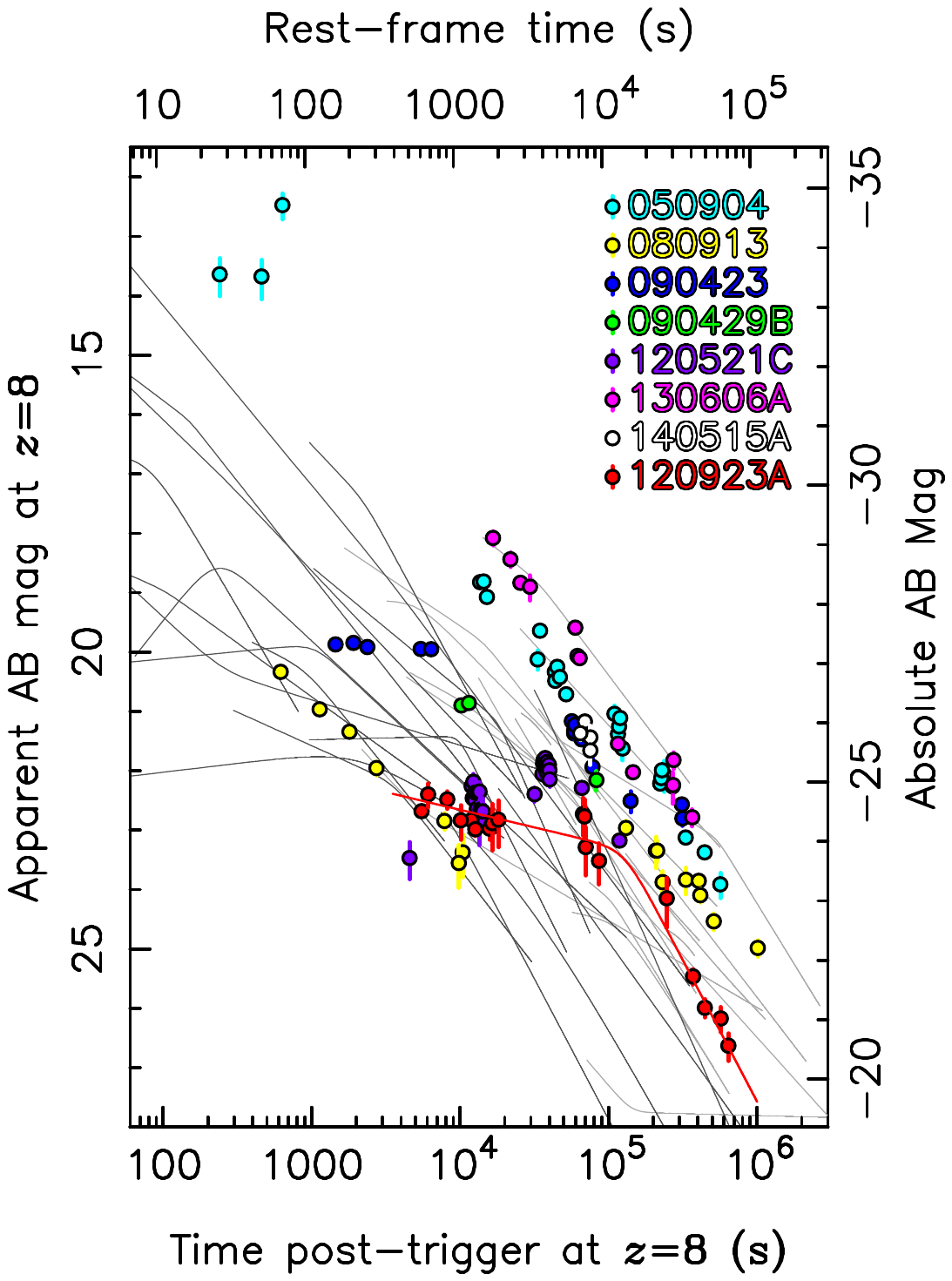}
\caption{Infrared light curves of high redshift GRBs\,120923A, 130606A \citep{Hartoog15},
120521C \citep{Laskar14},
050904 \citep{Boer06,Haislip06}, 
080913 \citep{Greiner09}, 090423 \citep{Tanvir09} and 090429B \citep{Cucchiara11}
as they would appear at redshift $z=8$ and also (alternative axes) rest-frame time and absolute 
magnitude. In grey are the broken power-law fits to light curves of a sample of low redshift afterglows from \citet{Kann10}, also transformed to 
the same redshift.
\label{fig:nir}}
\end{figure}

We present the GRB\,120923A composite NIR light curve formed by the $JHK$ and F140W photometry in 
Figure~\ref{fig:lc}, scaling the $JHK$ bands by small factors to match the F140W band. A fit 
to this overall light curve of a broken power-law model yields a shallow initial slope, $\alpha_1\approx-0.25$, breaking at 
$\approx35$ hours to a steep decay with $\alpha_2\approx-1.9$ ($\chi^2$/dof\,$=12.1/13$).
We compare the light curve to other high-redshift GRBs in Figure~\ref{fig:nir}. The 
afterglow of GRB\,120923A is comparatively faint, and could easily have escaped detection in other 
circumstances; i.e. we were lucky in being able to observe the afterglow with an 8-m telescope in 
excellent seeing within 2 hours, and to continue observations for several hours before the source 
set. Since the peak flux density of the afterglow SED is directly proportional to the blastwave 
kinetic energy \citep{gs02}, the relative faintness may result at least in part from a comparatively low value of 
$\EK$. We investigate this further via multi-wavelength modelling in Section \ref{sec:model}.

It is interesting to note that the SED of this afterglow is comparatively blue ($\beta=-0.17^{+0.34}_{-0.25}$ at 
$\approx0.14$~d; Figure \ref{fig:sed} and Section \ref{sec:photoz}), consistent with little line-of-sight dust extinction in 
the host, as has generally been found for other afterglows of the high-$z$ GRBs 
\citep{Zafar11,Laskar14}. This may reflect the limited time to build up dust, particularly 
in the small galaxies that are likely dominating the total star formation budget at $z>6$ 
\citep[although see][for an example of substantial dust in a galaxy at $z\approx7.5$]{Watson15}. 
Of course, there is also an observational bias against discovering dusty afterglows at high redshift.
We consider the quantitative constraints on dust extinction to GRB\,120923A in Section \ref{sec:multi}.

\section{Multi-wavelength Modelling}
\label{sec:model}
\subsection{Synchrotron Model}
We now interpret the multi-wavelength observations of GRB\,120923A in the context of the standard synchrotron 
model, in which the afterglow radiation arises from the blastwave shock set up by the expanding 
relativistic GRB ejecta interacting with their circumburst environment. 
This model assumes an idealised jet and ambient medium, but is appropriate given our limited
sampling of the evolving SED.
The resulting radiation 
is expected to exhibit characteristic power law spectral segments connected at `break frequencies', namely 
the synchrotron cooling frequency (\nuc), the characteristic synchrotron frequency (\numax), and 
the self-absorption frequency (\nua). The location of these frequencies depend on the physical 
parameters: the isotropic-equivalent kinetic energy (\EKiso), the circumburst density (\dens, or the 
normalised mass-loss rate in a wind-like environment, \Astar), the fraction of the blastwave energy 
imparted to non-thermal electrons (\epse), and the fraction converted into post-shock magnetic 
energy density (\epsb).

The different possible orderings of the break frequencies (e.g., $\numax < \nuc$: `slow cooling', 
and $\nuc < \numax$: `fast cooling') then give rise to five possible afterglow SED shapes 
\citep{gs02}. As the radius and Lorentz factor of the blastwave change with time, these break 
frequencies evolve and the afterglow SED may transition between these different spectral shapes. To 
preserve smooth light curves when break frequencies cross, we use the weighting schemes described 
in \cite{Laskar14} to compute the afterglow SED as a function of time. As in Section~\ref{sec:photoz}, we adopt the SMC 
extinction curve \citep{pei92} to model the extinction in the host galaxy, \AV, and include the 
possibility of an achromatic `jet-break' in the afterglow light curves due to spreading and 
edge-effects expected for a collimated outflow. To efficiently sample the available parameter 
space, we carry out an MCMC analysis using \texttt{emcee}. The details of our modelling scheme and MCMC 
implementation are described in \cite{Laskar14}.

\subsection{Basic Considerations}
\label{sec:basic_considerations}
The lack of any flattening in the late-time NIR light curve (Fig.~\ref{fig:nir}) indicates that any host
contamination of the afterglow photometry is small, and we assume it to  be negligible in what follows.
The $J$-band light curve declines as $\alpha_{\rm J} = -0.23\pm0.04$ between 0.06\,days and 
0.2\,days. In the synchrotron model, such a shallow decline in the NIR is only possible if $\nuc \lesssim 
\nuNIR < \numax$, where the lightcurves decline as $t^{-1/4}$ regardless of the density profile of 
the circumburst environment. This suggests that the afterglow radiation is in the fast cooling 
regime, and that the NIR bands are on segment F of \cite{gs02}.
 In this scenario, we would expect the light curve to steepen to $\alpha = (2-3p)/4$ when 
$\numax\propto t^{-3/2}$ passes through the NIR band, where $p$ is the power-law index of the
electron energy distribution. Alternatively, in the particular case of a 
wind environment, $\nuc \propto t^{1/2}$ may pass through the NIR band first, and the NIR decay 
rate would steepen to $\alpha = -2/3$.

The steepness of the late decay ($\alpha_{\rm J} \approx -1.8\pm0.3$ between $4.3$\,days and $20$\,days),
and marked change from the early behaviour, provides evidence that a jet break occurred at $\lesssim4.3$\,days,
and also suggests the passage of $\numax$ through the NIR band between  $0.2$ and $4.3$\,days, and indicates a uniform 
(ISM-like) environment. If we take the power law within this window, $\alpha_{\rm J} = -1.2\pm0.2$, as indicative
of the slope after $\numax$ passage, but before the jet break, then using $\alpha_{\rm J} = (2-3p)/4$ yields $p=2.3\pm0.3$.

The XRT PC-mode light curve is well-fit with a broken power law, with an initial flat segment, $\alpha_{\rm X, 1} = 0.0\pm0.2$, breaking 
into\footnote{The smoothness ($y$ in \citealt{Laskar14}; equivalent to $s$  in \citealt{gs02}) of the break is poorly  constrained, and we fix it to $y=3$.} 
$\alpha_{\rm X, 2} = -1.32\pm0.05$ at $t_{\rm b} = (6.3\pm1.2)\times10^{-3}$\,days. 
The initial flat portion of the X-ray light curve may be due to the 
X-rays being on the same segment (F) of the synchrotron SED as the NIR data. In this case, the 
X-ray decline rate is also expected to be $t^{-1/4}$. The break in the X-ray light curve would then 
correspond to the passage of $\numax$ through the X-ray band. Since the X-ray band spans an order 
of magnitude in energy, while $\numax$ evolves as $t^{-3/2}$, we would expect the break to be 
smoothed out over a factor of $\approx10^{-2/3}\approx5$ in time. If we assumed $\numax \approx \nu_{\rm 
J}\approx 2.2\times10^{14}$\,Hz at $\approx0.2$\,days, we would expect $\numax\approx 1$\,keV at 
$\approx2\times10^{-3}$\,days. This is consistent with the observed steepening in the X-ray lightcurve 
at $\approx6\times10^{-3}$\,days. In this model, the post-break decline rate of $\alpha_{\rm X} = 
-1.32\pm0.05$ yields $p=2.43\pm0.07$. The different decline rates in the X-ray and NIR bands 
between 0.06\,days and 0.2\,days suggests that these bands are on different segments of the afterglow 
synchrotron spectrum, consistent with the spectral ordering, $\nuc \lesssim \nuNIR < \numax < \nuX$ during 
this period.

The spectral index in segment F is expected to be $\beta = -0.5$, independent of $p$. When the 
X-rays are on this segment, we would expect $\Gamma_{\rm X} = 1-\beta_{\rm X} = 1.5$, which is 
consistent with the value of $\Gamma_{\rm X} = 1.61\pm0.14$ derived in Section \ref{sec:Swift}. 
We would also expect spectral evolution from $\Gamma_{\rm X} = 1.5$ to $\Gamma_{\rm X} = 1+p/2$ 
after $\numax$ crosses the X-ray band. Unfortunately, paucity of data following the orbital gap 
in the X-ray light curve precludes confirmation of this behaviour.

Interpolating the X-ray lightcurve using the best fit broken power law model, we find 
a flux density of $(3.5\pm0.4)\times10^{-2}$\,$\mu$Jy at the time of the Gemini/NIRI $J$-band observation at 
$0.064$\,days. The spectral index between the NIR and X-ray band is then $\beta_{\rm 
NIR-X}\approx-0.65\pm0.01$. This is significantly different from $-0.5$, suggesting that at least 
one spectral break frequency lies between the NIR and X-ray band. 
Assuming a spectral slope of 
$\beta=-0.5$ and $\beta=-p/2=-1.22$ below and above this break, respectively, and using the 
measured $J$-band flux density and extrapolated X-ray flux density, we can locate the break to be 
at $\approx5.9\times10^{16}$\,Hz at this epoch.
However, extrapolating 
$\numax\propto t^{-3/2}$ from $\approx2.2\times10^{14}$\,Hz at $\approx0.2$\,days to $0.064$\,days
yields 
$\numax \approx1.2\times10^{15}$\,Hz, which is more than an order of 
magnitude lower. 
Here we have neglected the possibility of dust extinction in the host galaxy.  
A small amount of extinction would lead us to overestimate $\beta_{\rm NIR-X}$
and hence overestimate $\numax$, and so in principle alleviate this discrepancy.
However, we estimate that requiring $\beta_{\rm NIR-X}\approx\beta_{\rm X}$ 
would necessitate $A_{\rm V}\gtrsim0.2$, which is disfavored from our analysis of the NIR photometry in Section~\ref{sec:photoz}.
We return to the question of host extinction in Section~\ref{sec:multi}.

In the synchrotron model, we can use the observed X-ray flux density at 1\,keV at a time that is 
dominated by afterglow radiation to estimate the burst kinetic energy \citep[e.g.][]{Granot06}. This requires the X-ray band 
to be located above the peak and cooling frequencies. Since $\nuc,\numax<\nuX$ after the break in 
the X-ray light curve, this condition is satisfied. We use the last point preceding the \Swift\ 
orbital gap with $f_{\rm X}\approx0.22\,\mu$Jy at 
$\approx0.36$\,hr, 
together with fiducial values of $p=2.2$ \citep[e.g.,][]{curran10} and $\epse=\epsb=1/3$ to estimate $\EKiso\approx3\times10^{51}$\,erg. 
We verify this result in the next section.

To summarise, the X-ray light curve, X-ray spectrum, and NIR $J$-band light curve suggest that the 
observed synchrotron radiation from the blastwave shock is in the fast cooling regime with $\nuc \lesssim 
\nuNIR < \numax < \nuX$ at $\approx 0.2$\,days; $\numax$ passes through the X-ray band at about a few 
$\times 10^{-3}$\,days, and through the $J$-band between $\approx0.2$\,days and $4.3$\,days, while the steep decline 
in the \HST\ F140W data indicates a jet break before $\approx 4.3$\,days.

\begin{figure}
\includegraphics[width=95mm]{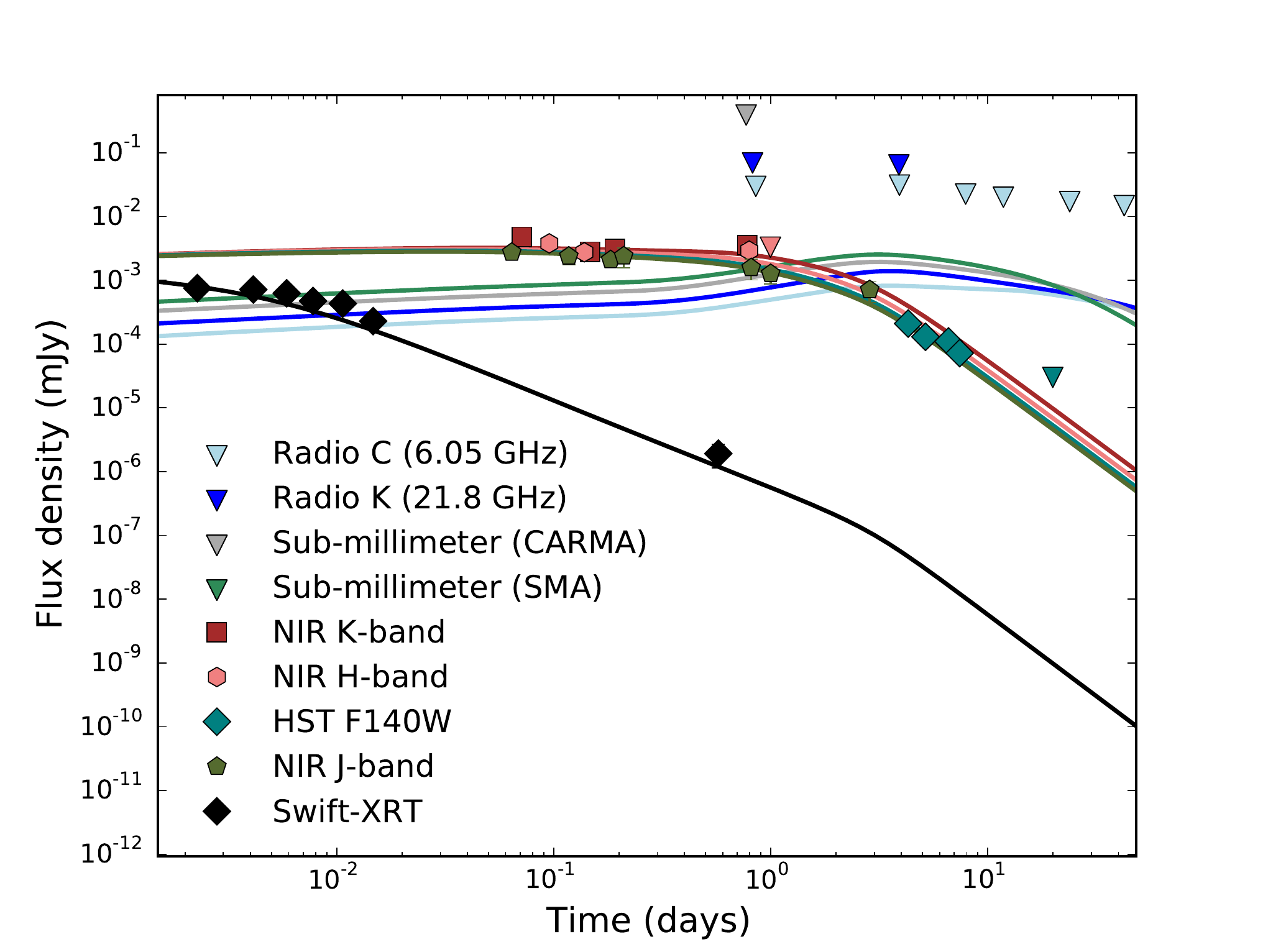}
\caption{X-ray, NIR, and radio light curves of GRB\,120923A, with the best-fit model (solid lines) 
fitted simultaneously to all the available multi-wavelength data.
\label{fig:multi}}
\end{figure}

\begin{figure}
\includegraphics[width=90mm]{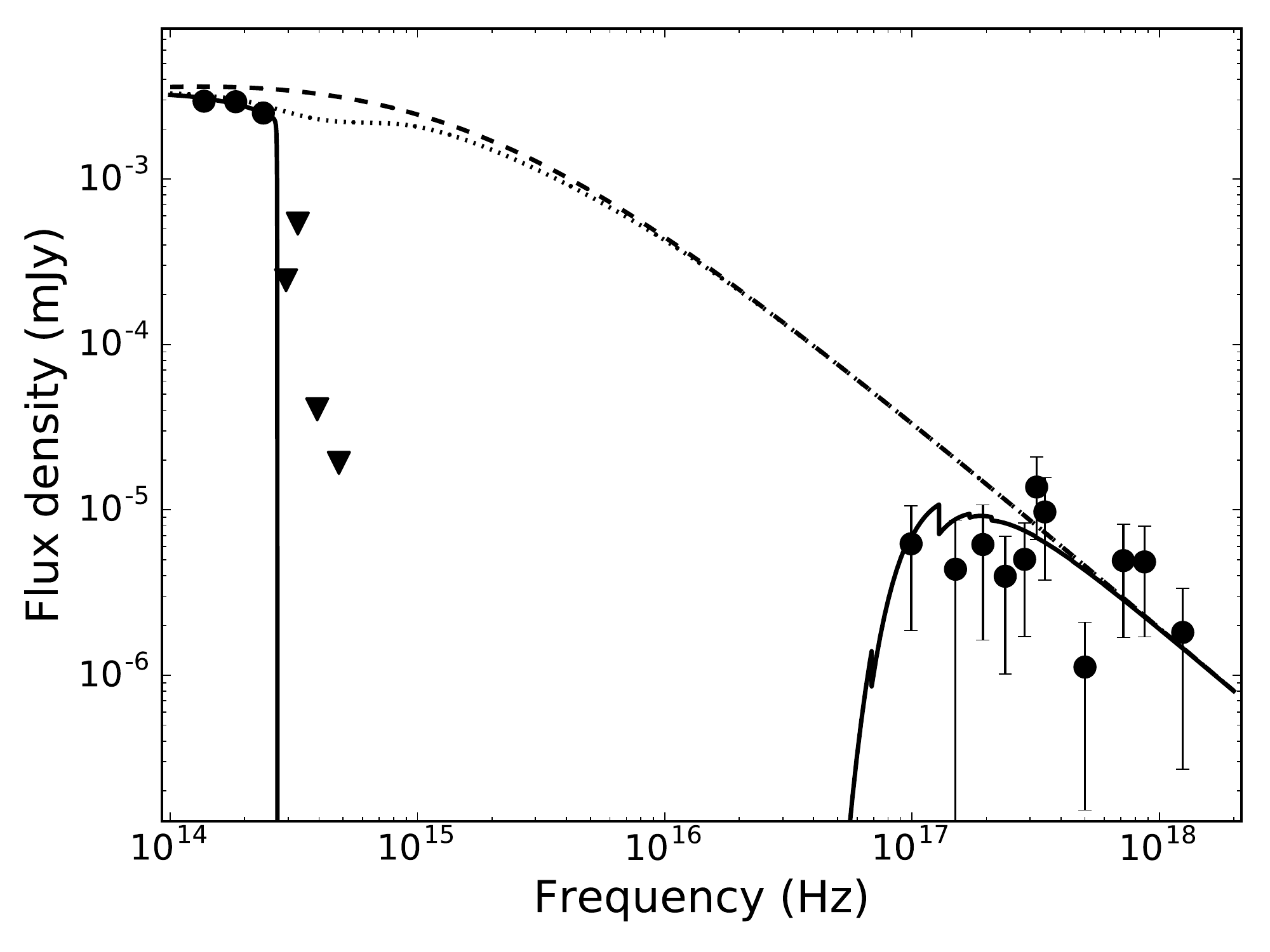}
\caption{NIR to X-ray SED of the GRB\,120923A afterglow at 0.14\,days together with the best-fit model 
(solid). The NIR observations are the same points used in Figure~\ref{fig:sed} and have been 
corrected for Galactic extinction. The X-ray SED is computed from PC-mode data after 
$1.1\times10^{-2}$\,days and interpolated to 0.14\,days using the X-ray light curve decline rate of 
$\alpha_{\rm X,2}\approx-1.32$ (Section \ref{sec:basic_considerations}). The intrinsic afterglow 
synchrotron SED (without IGM absorption or host extinction) is shown as the dashed black curve, 
while the dotted curve is the intrinsic SED with $\AV=0.06$\,mag of extinction in the host galaxy. 
Although a small amount of extinction is present in the best fit model, our MCMC analysis indicates 
that overall evidence for extinction in the host galaxy along the line of sight is marginal.
\label{fig:xrt_sed}}
\end{figure}

\subsection{Multi-wavelength Model for GRB\,120923A}
\label{sec:multi}
We now describe the full multi-wavelength modelling of all available afterglow data for GRB\,120923A 
using the techniques presented in \cite{Laskar14}, which include accommodating upper limits. 
We adopted weak, flat priors based on plausible ranges: $2.01<p<3.45$, 
$\epse,\epsb<1/3$, $-10<\log(\dens/{\rm cm^{-3}})<10$, $-4<\log(\EKiso/10^{52}\,{\rm erg})<2.7$, $\AB < 20$\,mag, 
and $-5<\log(\tjet/{\rm d})<5$. 
For the sake of generality, we did not fix the redshift, but 
based on our analysis of the NIR spectral energy distribution (Section \ref{sec:specz}) restricted the redshift range to $7.0<z<8.5$. 

We find that an ISM-like model with a jet break adequately explains the full set of afterglow 
observations. Our highest likelihood model (Figure~\ref{fig:multi}) has the parameters 
$p\approx2.5$, $z\approx8.1$, $\epse\approx0.33$,  $\epsb\approx0.32$, 
$\dens\approx4.0\times10^{-2}\,{\rm cm^{-3}}$, and $\EKiso\approx2.9\times10^{51}$\,erg, with a jet break at 
$\tjet\approx3.0$\,days ($\chi^2$/dof $ = 1.1$). 
We note that the derived value 
of \EKiso\ confirms the estimate made using the X-ray data (Section \ref{sec:basic_considerations}).
An implication of this is a very high value for the radiative efficiency $\eta=\Egammaiso/(\Egammaiso+\EKiso)\approx0.92$.
Such high values for $\eta$ have also been inferred for the prompt emission of some other GRBs \citep[e.g.][]{Zhang07}, although they 
remain a challenge to explain theoretically.

Thus the redshift derived by this approach is completely consistent with the photometric redshift found in Section~\ref{sec:photoz}.
This model requires a small amount of extinction 
in the host galaxy, $\AV\approx0.06$\,mag (Figure~\ref{fig:xrt_sed}). Using the relation,
\begin{equation}
 \theta_{\rm jet} = 0.1\left(\frac{\dens}{E_{\rm K,iso}/10^{52}}\right)^{1/8}
                       \left(\frac{\tjet/(1+z)}{6.2\,{\rm hr}}\right)^{3/8}
\end{equation}
for the jet opening angle \citep{sph99} we find $\thetajet\approx4.9$\,degrees and beaming-corrected 
kinetic energy, $\EK\approx1.1\times10^{49}$\,erg. 

The break frequencies are located 
at $\nuac\approx3\times10^{7}$\,Hz,  $\nusa\approx7\times10^{7}$\,Hz, $\nuc\approx6\times10^{14}$\,Hz, 
and $\numax\approx3\times10^{15}$\,Hz at 0.1\,days and the peak flux density is $\approx13\,\mu$Jy at 
$\nuc$ for the highest-likelihood model. In this model, \numax\ passes through 1\,keV at 
$\approx5\times10^{-3}$\,days, which is precisely the time of the observed break in the X-ray light 
curve, $t_{\rm b}=(6.3\pm1.2)\times10^{-3}$\,days (Section \ref{sec:basic_considerations}). The shallow-to-steep 
transition in the X-ray light curve is therefore consistent with the passage of \numax.

We note that the spectrum peaks at \nuc\ in the fast cooling 
regime, and the proximity of the cooling break to the NIR $J$-band at $\approx0.1$\,days results in a 
spectrum near the $J$-band that is flatter than $\nu^{-0.5}$. This explains the lower value of \numax\ 
(which lies between \nuc\ and \nuX) inferred from the NIR and X-ray light curves, compared 
with the value required in a broken power-law fit to match the NIR $J$-band and interpolated 
X-ray flux density at this time. The location of $\nuc \gtrsim \nu_{\rm J}$ at 0.1~d rules 
out the wind model, since in that case we would expect the NIR light curve to decline as $t^{-2/3}$ 
after $\approx 0.1$~d (Section \ref{sec:basic_considerations}). 
The passage of \numax\ through the NIR $J$-band occurs at $\approx0.5$\,days, at which
point both \nuc\ and \numax\ are below $\nu_{\rm J}$ and the light curve steepens to $\alpha\approx-1.4$, 
consistent with the limited observations at this time.
Finally, the best-fit model requires a jet break at $\approx3$\,days.

\begin{figure}
\centerline{\includegraphics[width=75mm]{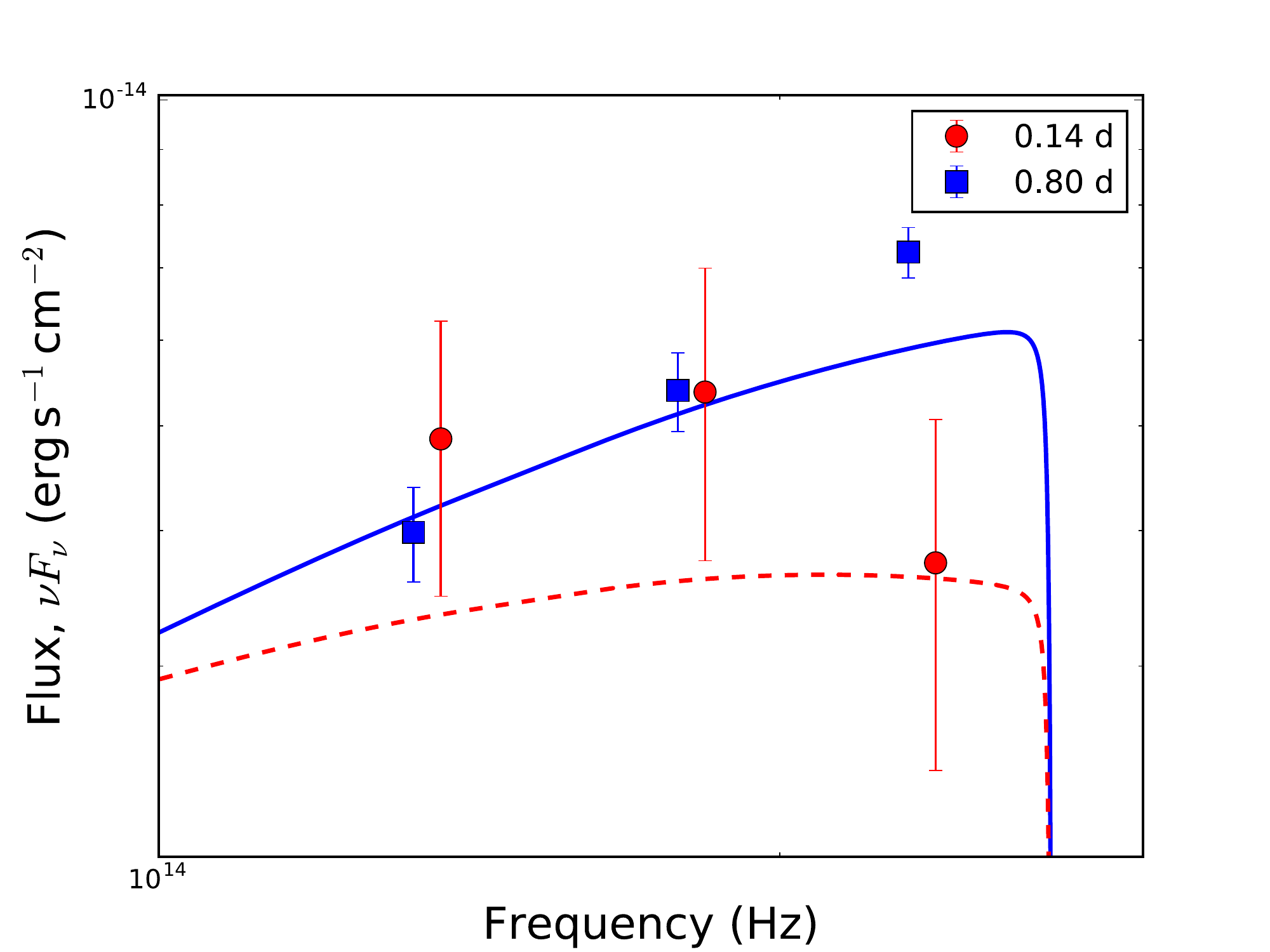}}
\caption{Spectral energy distribution ($\nu f_{\nu}$) of the {\em JHK} afterglow of GRB\,120923A at at 0.14\,days 
(red circles) and 0.8\,days (blue squares), together with the best fit multi-wavelength model (dashed and solid, respectively; Section~\ref{sec:multi}). 
The change in the slope of the SED is evident in the data, 
and is ascribed in the model to the transition of \numax\ and \nuc\ through the NIR bands after the transition to slow cooling at $\approx0.34$\,days.
\label{fig:ir_sed}}
\end{figure}

In broad agreement with the basic analysis presented in Section \ref{sec:basic_considerations}, the 
model afterglow SED remains in fast cooling until $\approx0.34$\,days   (approximately 1\,hr in the rest-frame). 
This conclusion is driven in the fit by the apparent change in NIR spectral slope between
the early data, particularly the Gemini photometry at $\sim0.14$\,days and the VLT epoch at $\sim0.8$\,days,
as illustrated in Figure~\ref{fig:ir_sed}.

From our MCMC simulations, we constrain the fitting parameters to $p=2.7^{+0.3}_{-0.2}$, 
$z=8.1^{+0.2}_{-0.3}$, $\epse=0.31^{+0.02}_{-0.04}$, $\epsb = 0.23^{+0.07}_{-0.11}$, 
$\dens = (4.1^{+2.2}_{-1.4})\times10^{-2}\,{\rm cm^{-3}}$, $\EKiso=(3.2^{+0.8}_{-0.5})\times10^{51}$\,erg, 
$\tjet = 3.4^{+1.1}_{-0.5}$\,days (68\% credible intervals), and $\AV\lesssim0.08$\,mag (90\% 
confidence upper limit). Thus, although the best-fit model has a small amount of extinction (Figure 
\ref{fig:xrt_sed}), our MCMC results indicate that evidence for dust along the line of sight is 
statistically marginal. 

\begin{figure*}
\begin{tabular}{ccc}
 \centering
 \includegraphics[width=0.30\textwidth]{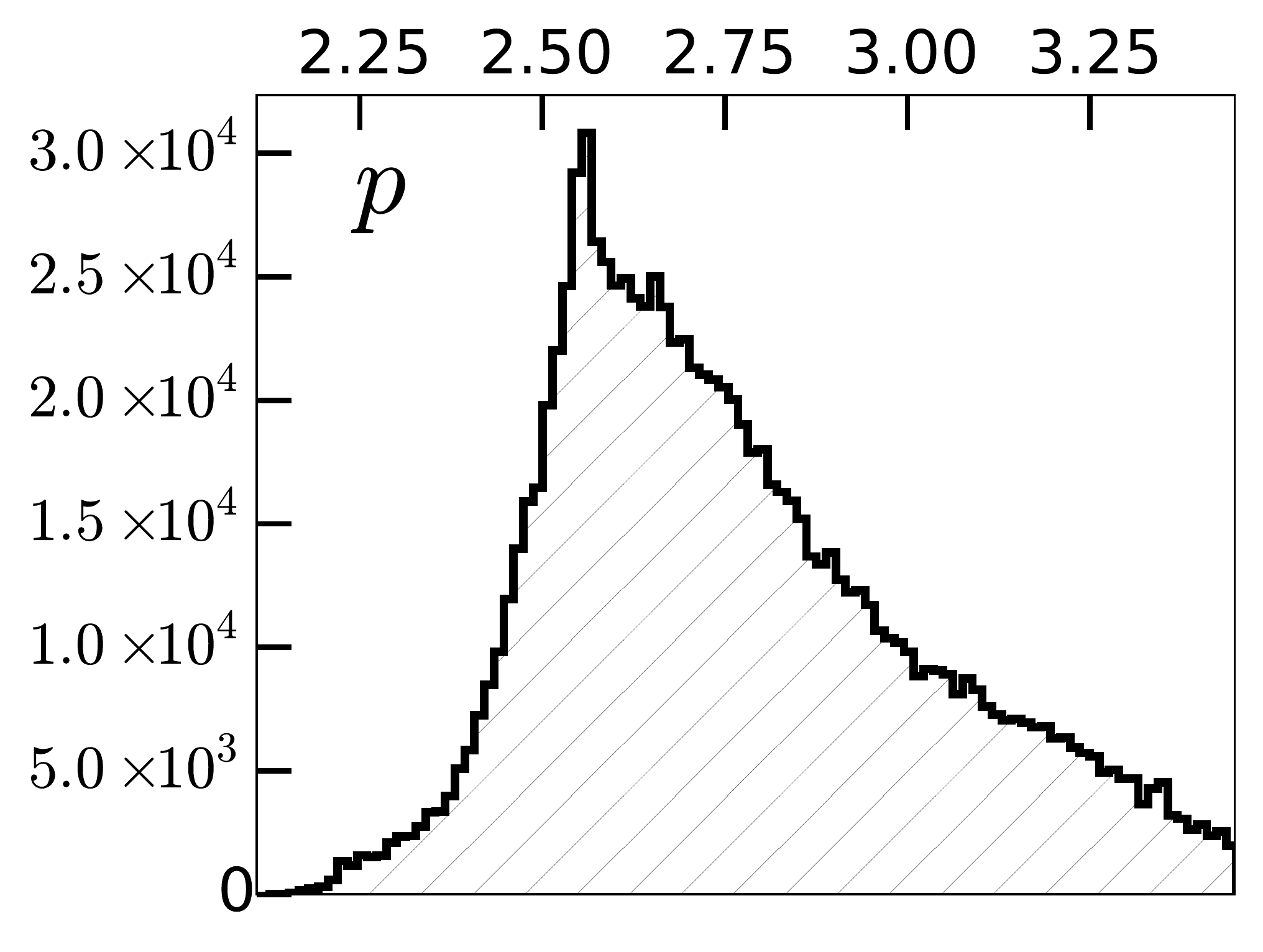} &
 \includegraphics[width=0.30\textwidth]{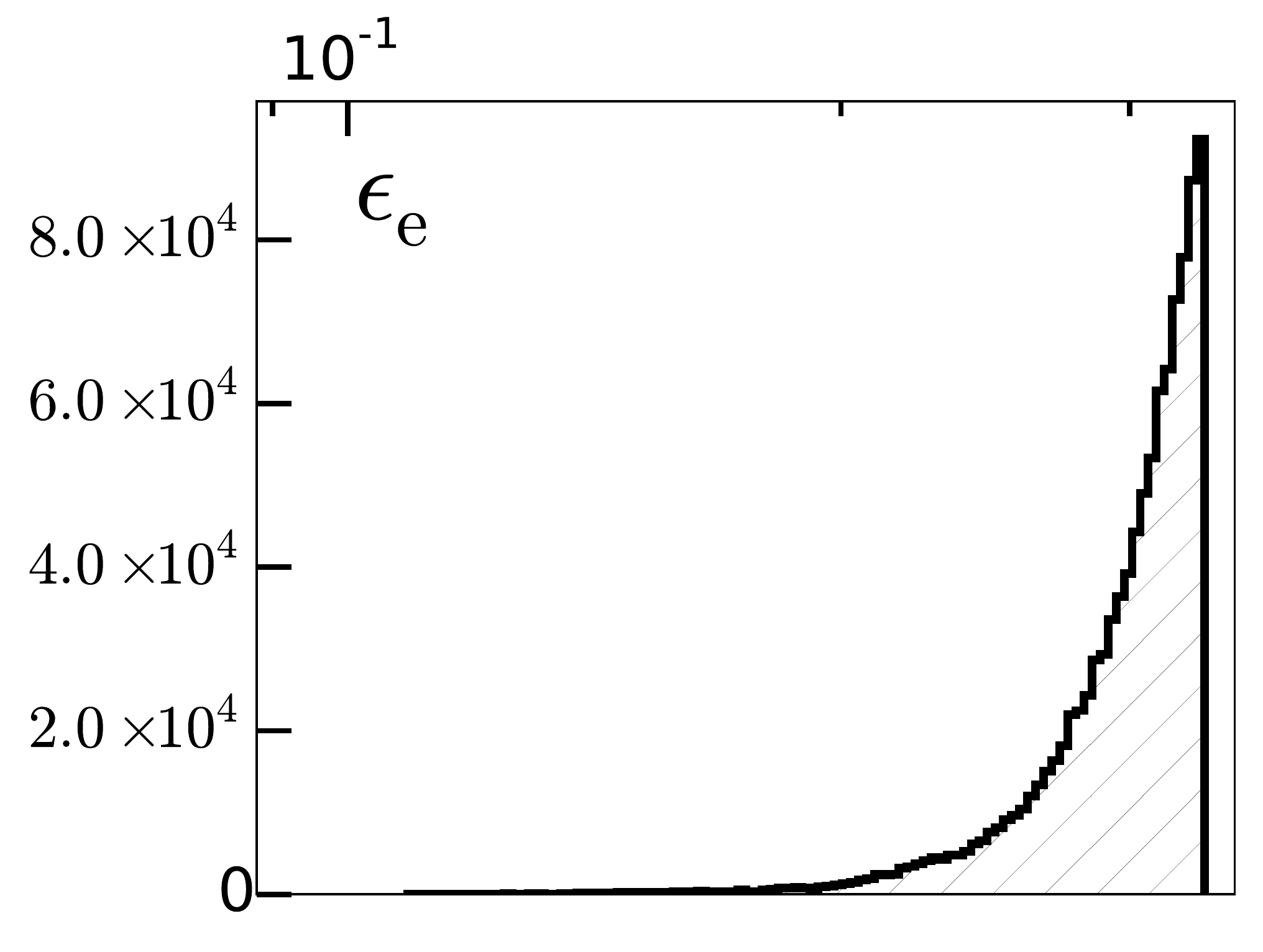} &
 \includegraphics[width=0.30\textwidth]{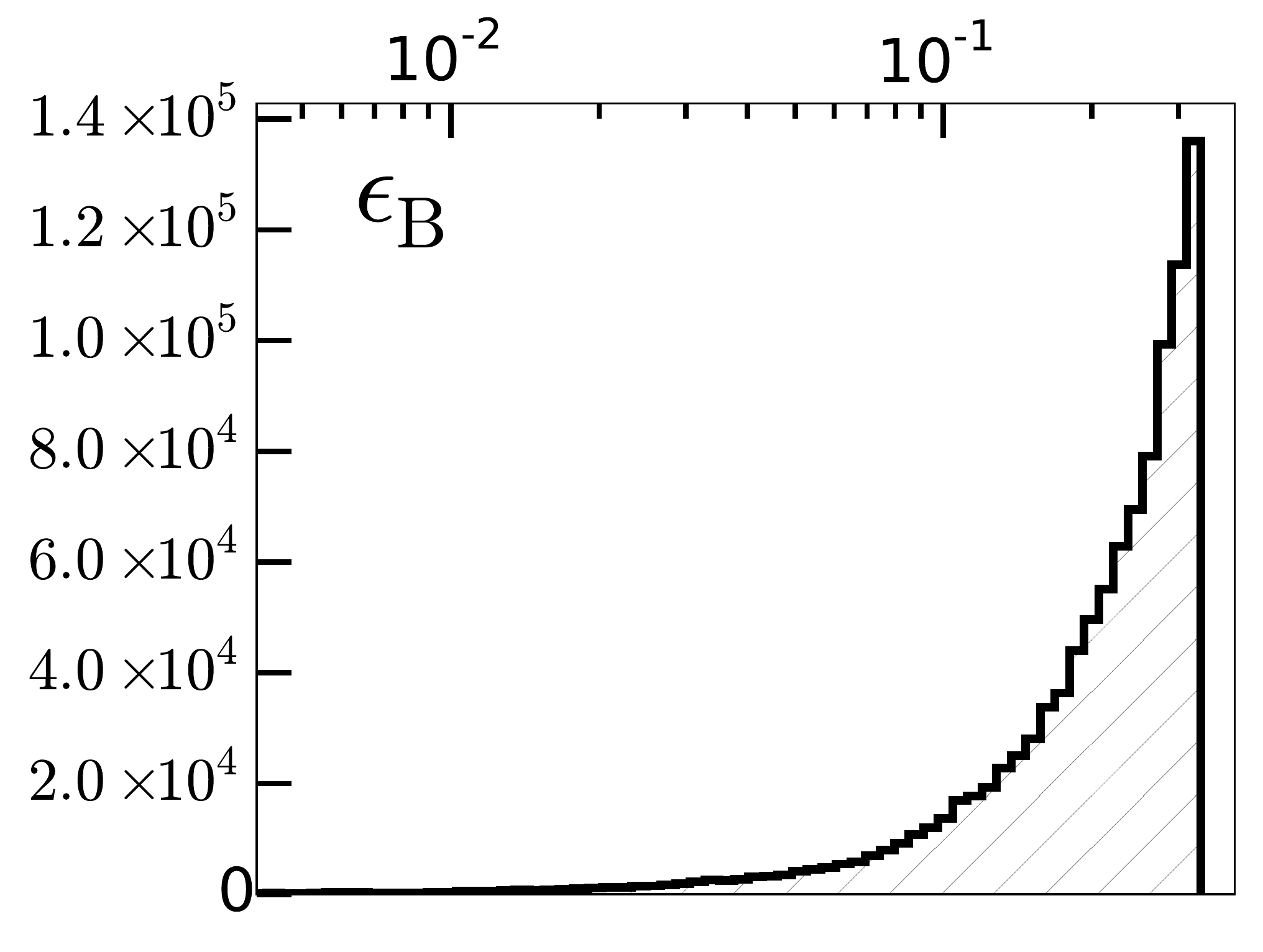} \\
 \includegraphics[width=0.30\textwidth]{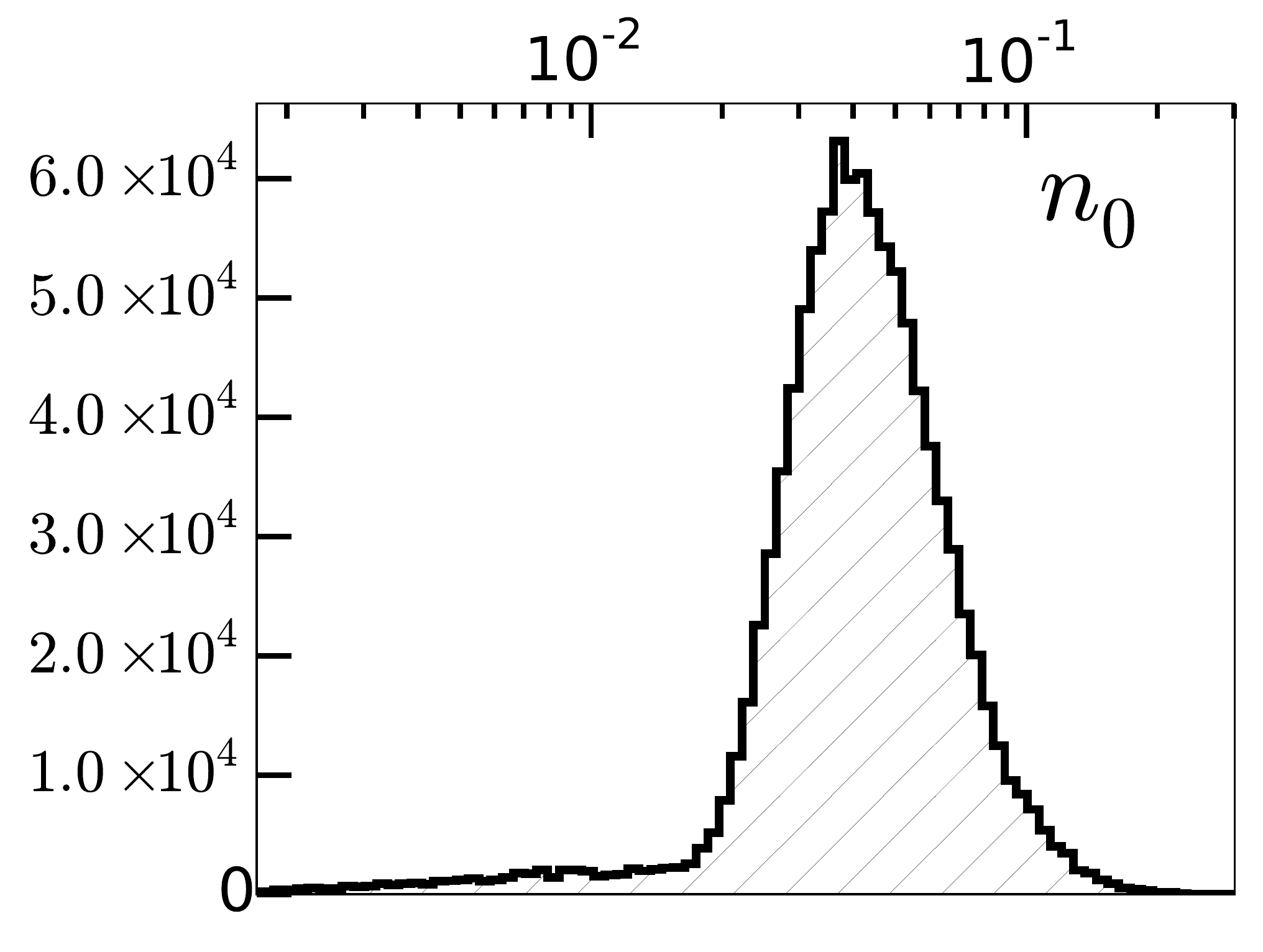} & 
 \includegraphics[width=0.30\textwidth]{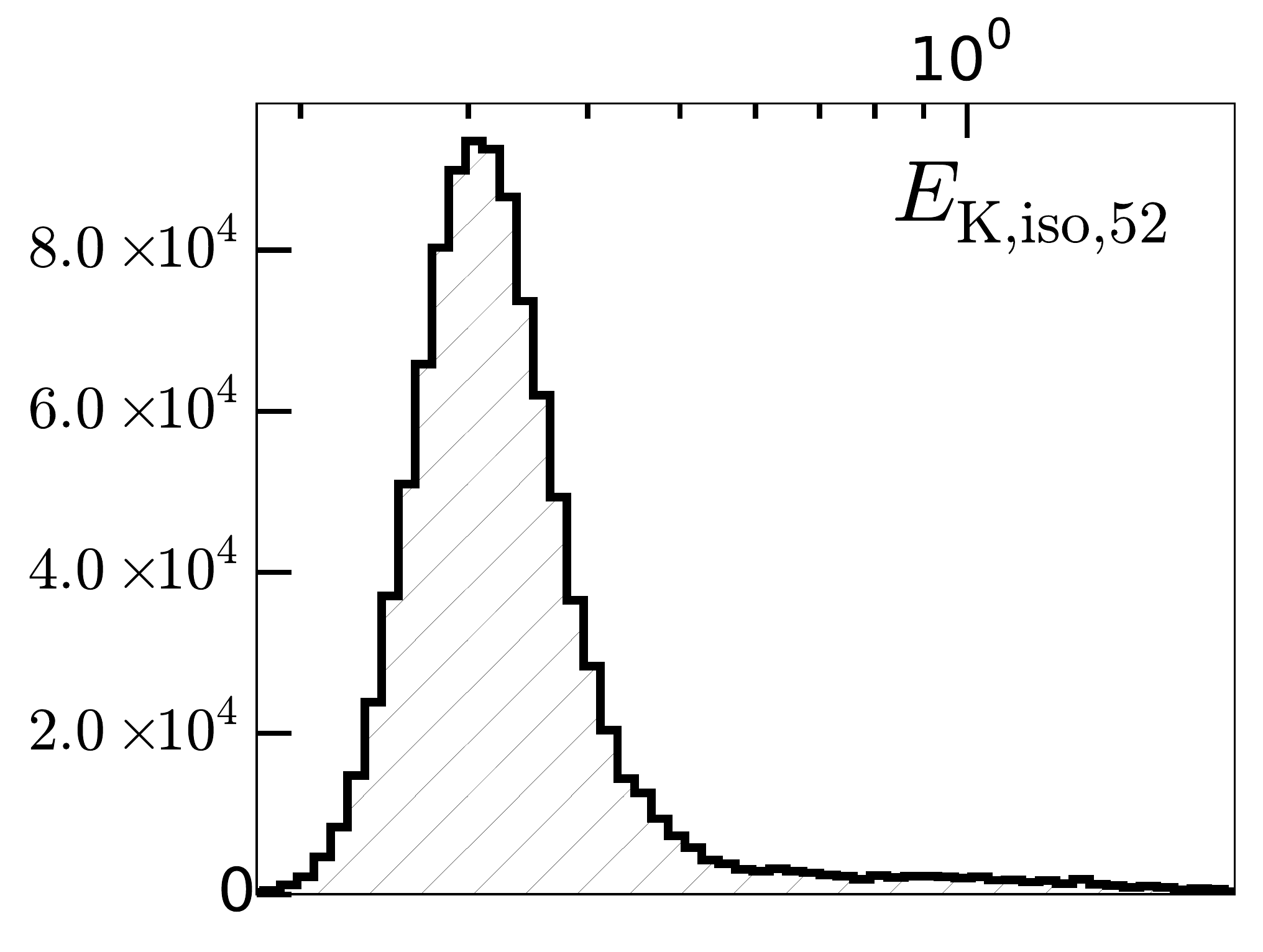} &
 \includegraphics[width=0.30\textwidth]{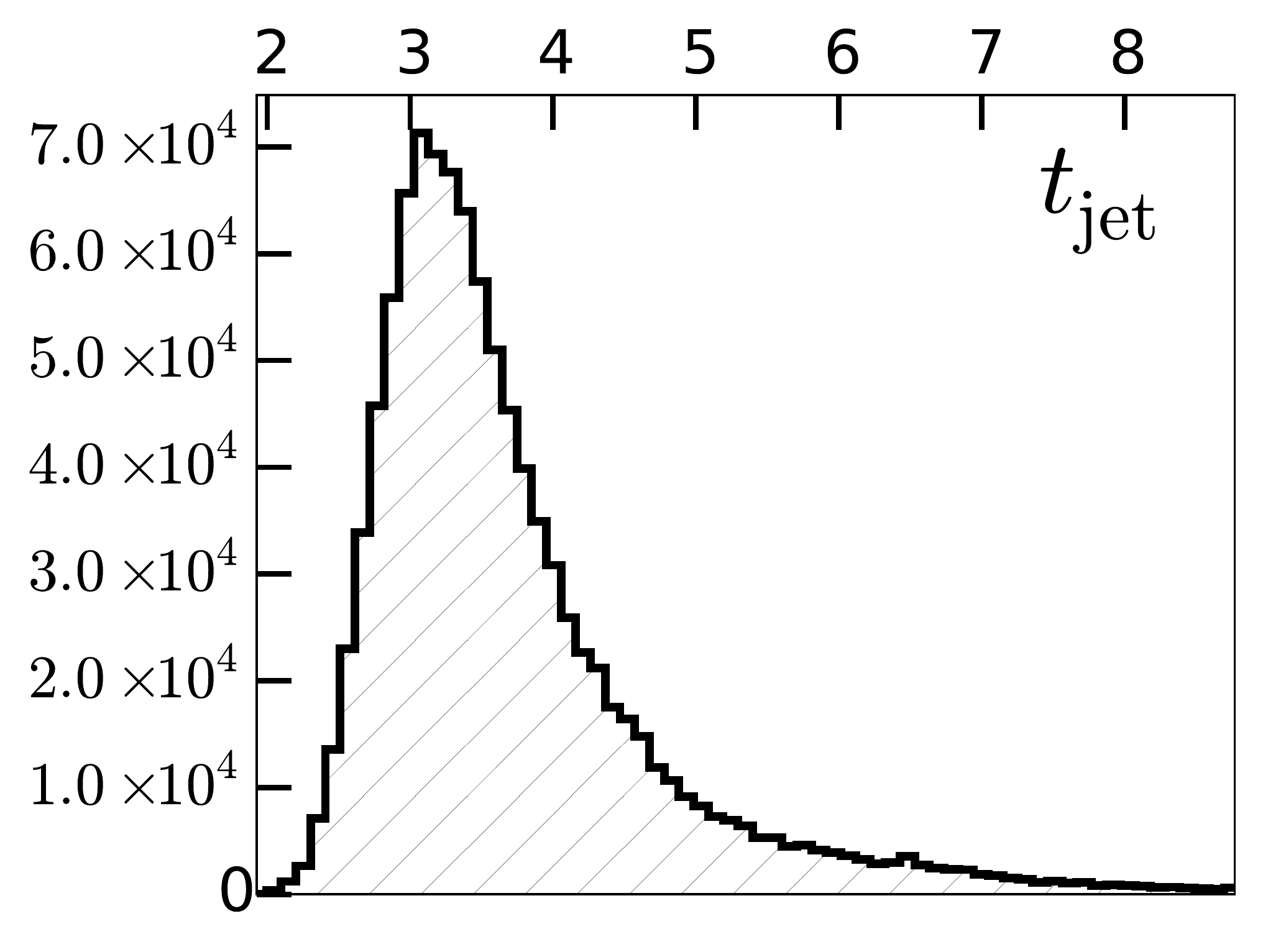} \\ 
 \includegraphics[width=0.30\textwidth]{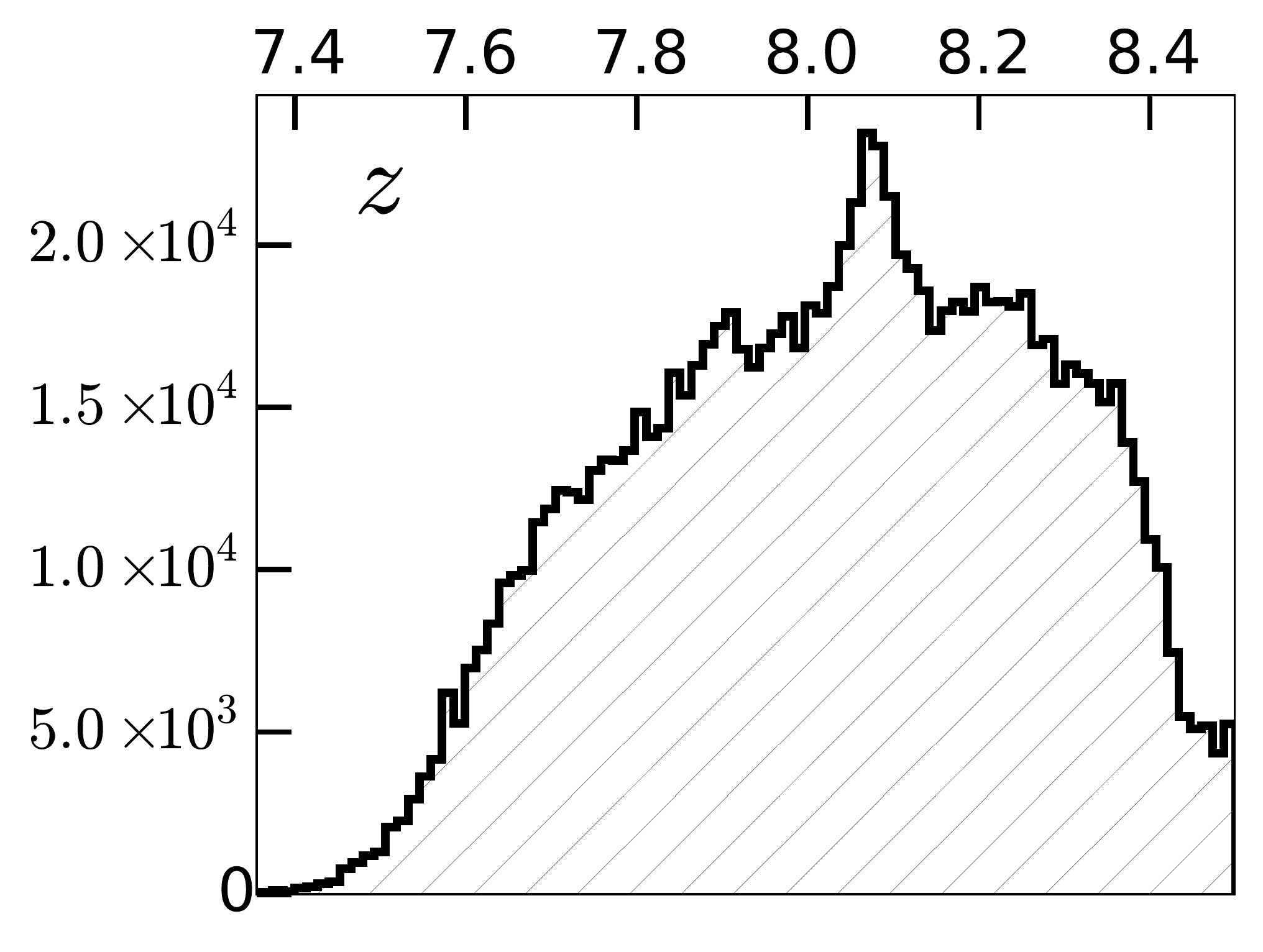} &
 \includegraphics[width=0.30\textwidth]{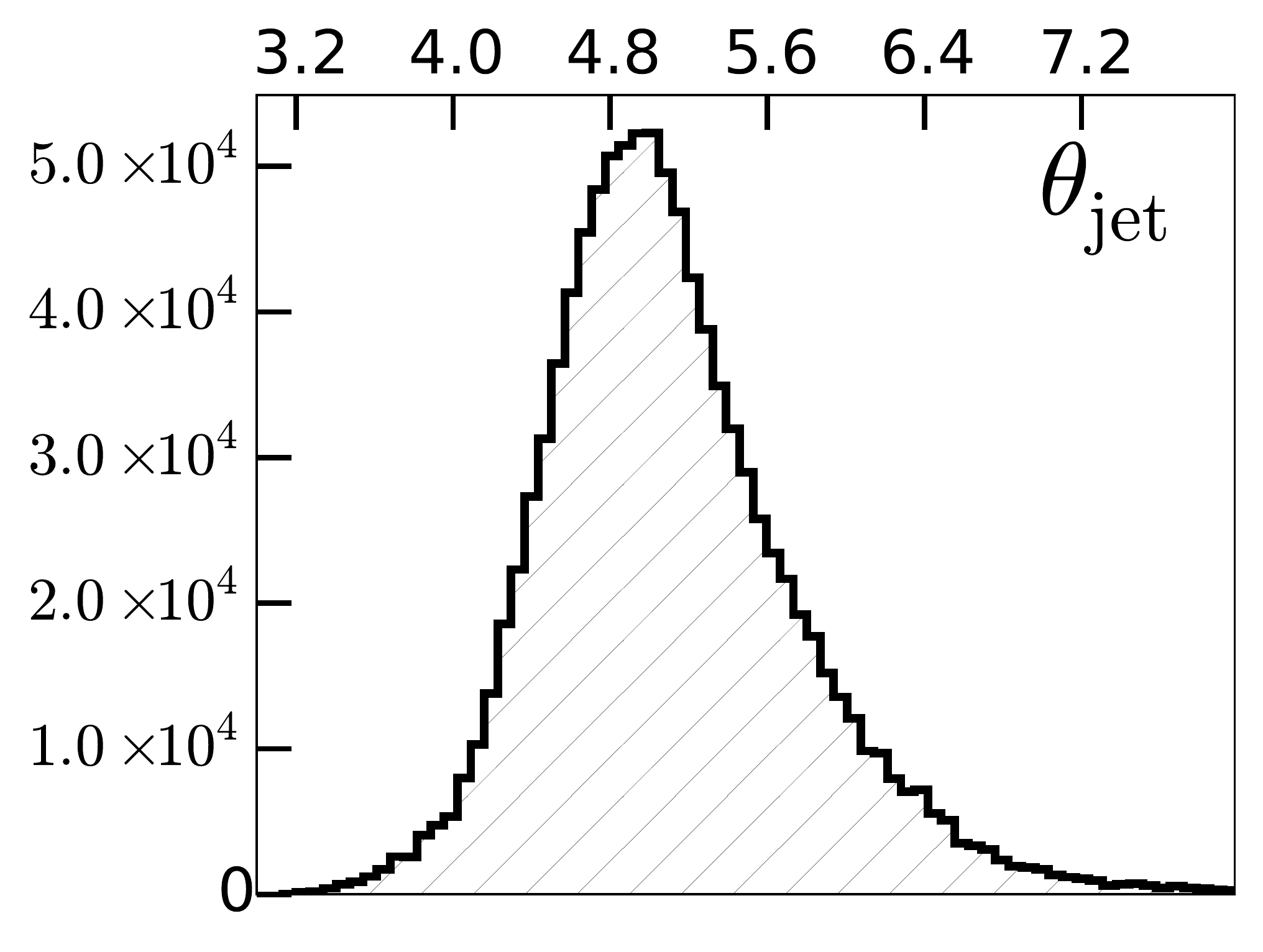}  &
 \includegraphics[width=0.30\textwidth]{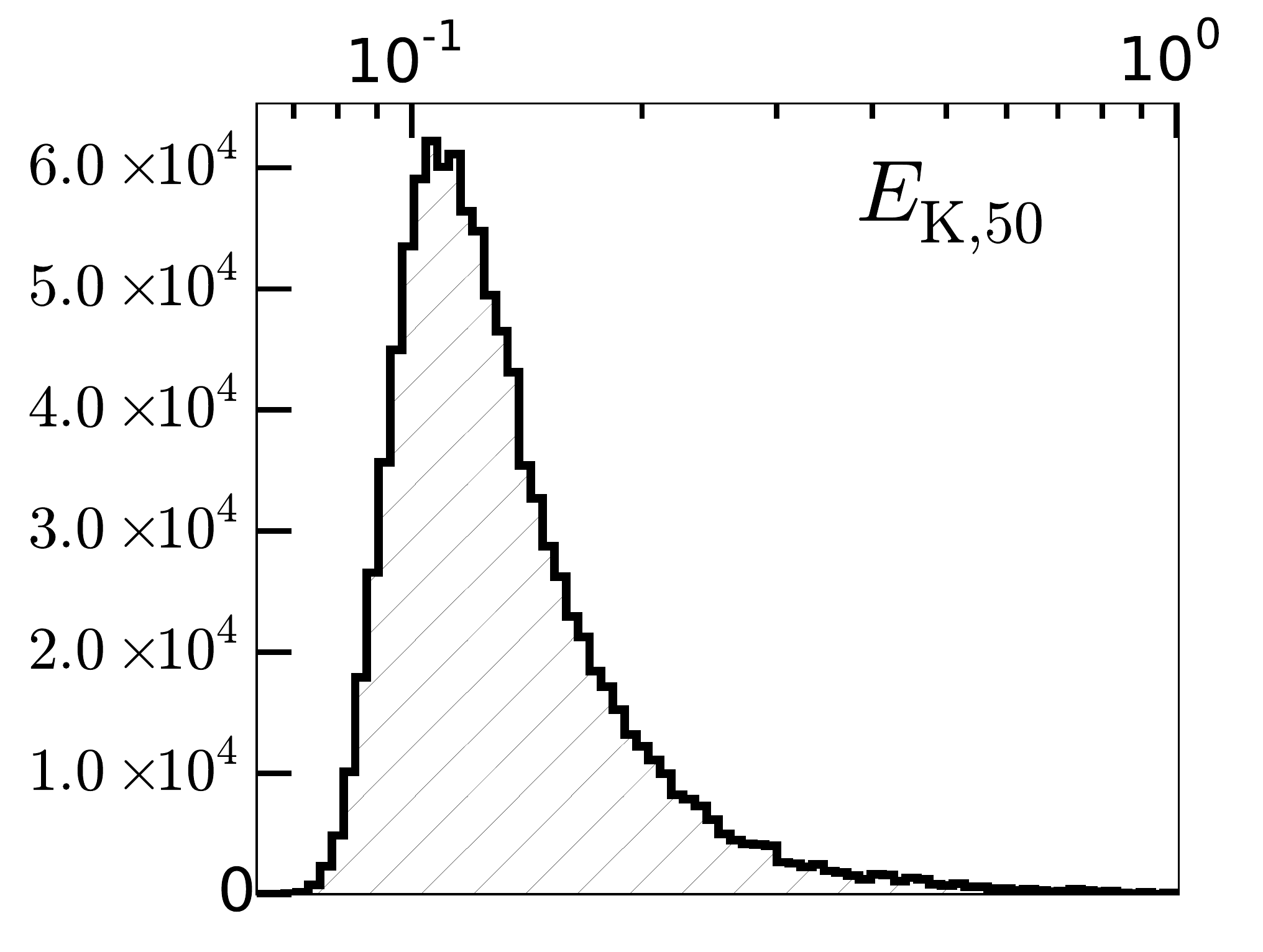} \\  
\end{tabular}
\caption{Posterior probability density functions for the physical parameters of GRB\,120923A in 
the ISM-like model from MCMC simulations. We have restricted $\epsilon_{\rm e} < 1/3$ and
$\epsilon_{\rm B} < 1/3$. Units of the quantities are: $n_0$ in cm$^{-3}$, $E_{\rm K,iso}$ and 
$E_{\rm K}$ in erg, $t_{\rm jet}$ in days, and $\theta_{\rm jet}$ in degrees.}
\label{fig:120923A_ISM_hists}
\end{figure*}

\begin{figure*}
\begin{tabular}{ccc}
\centering
 \includegraphics[width=0.30\textwidth]{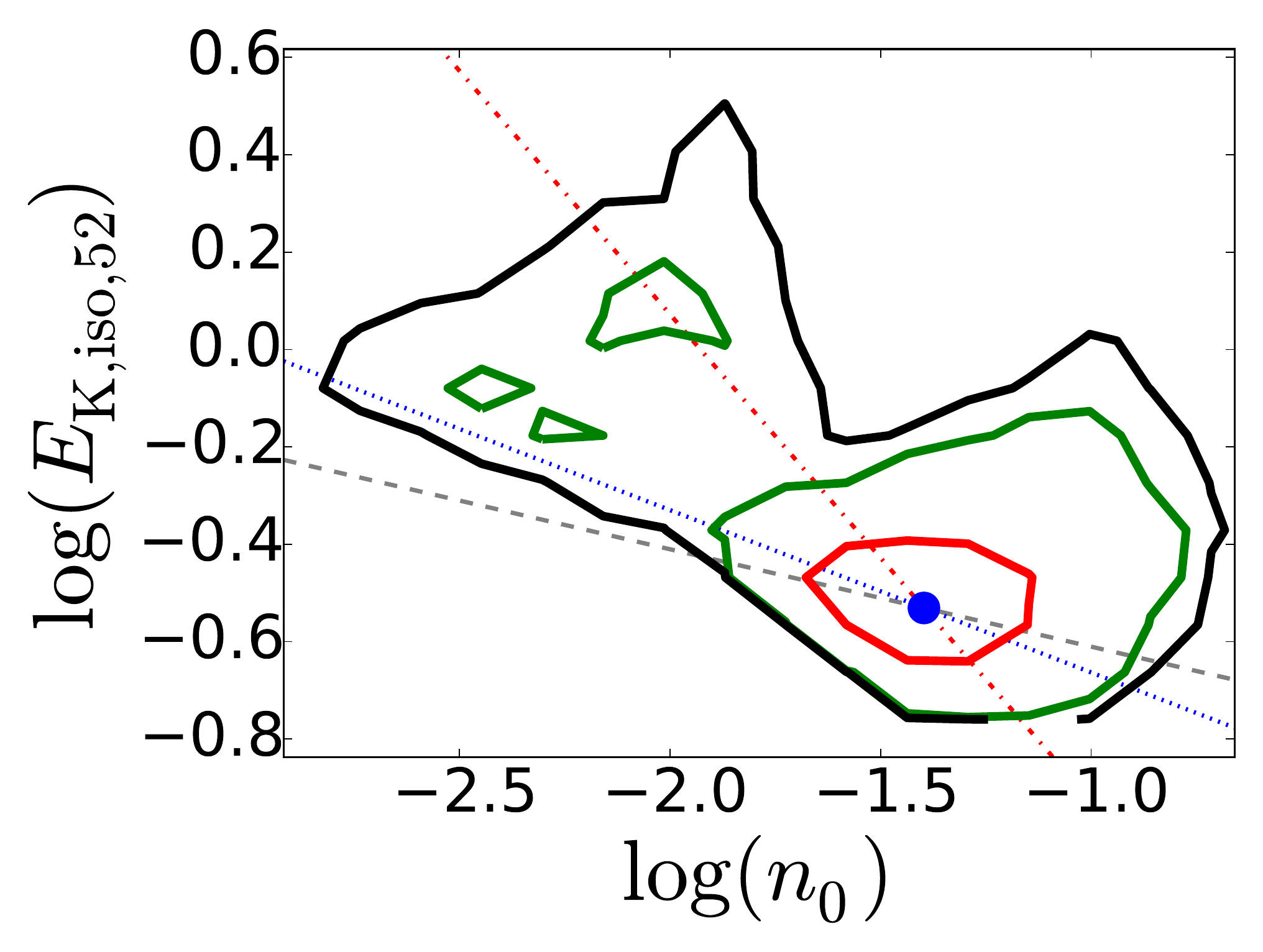} &
 \includegraphics[width=0.30\textwidth]{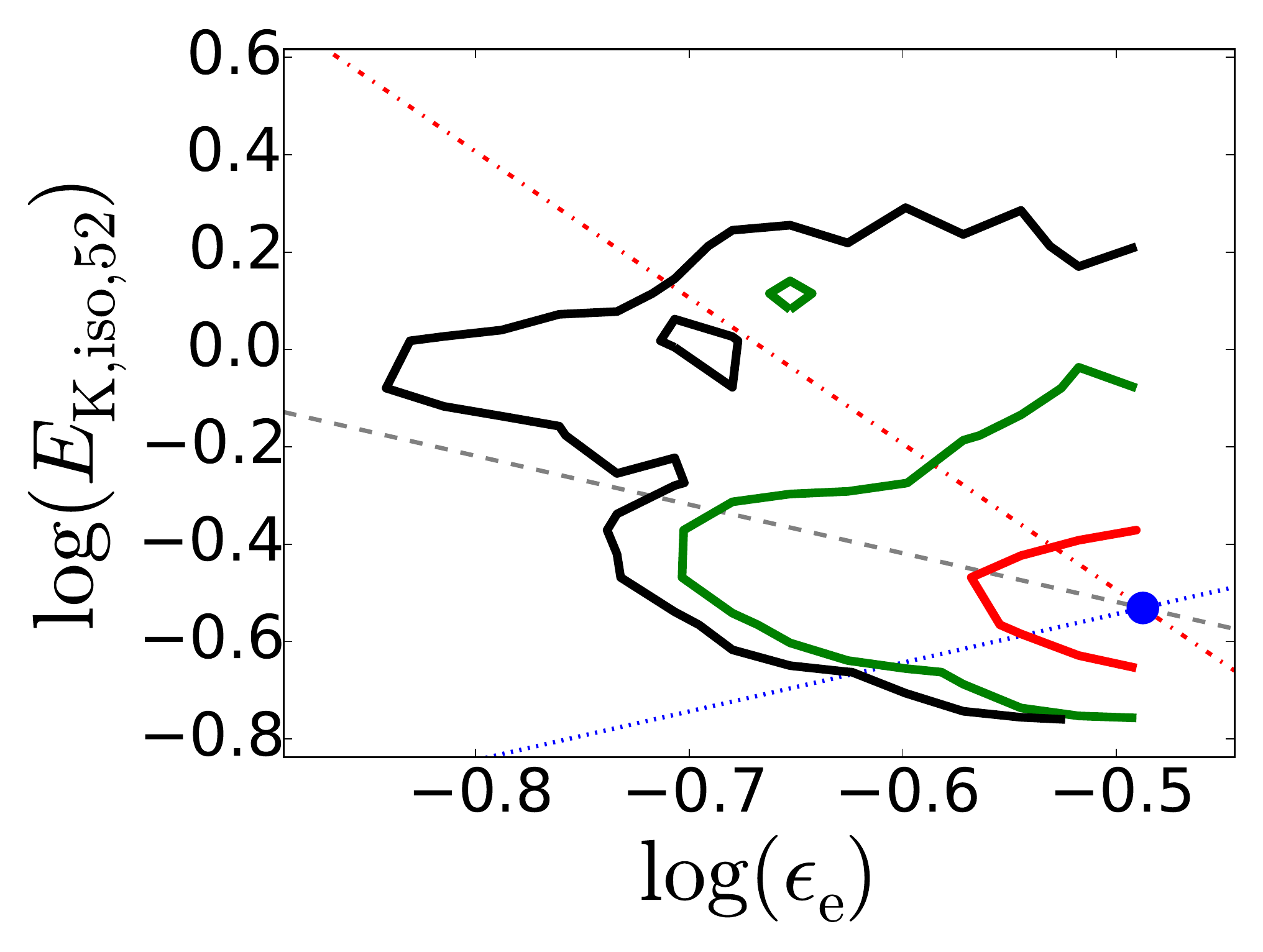} &
 \includegraphics[width=0.30\textwidth]{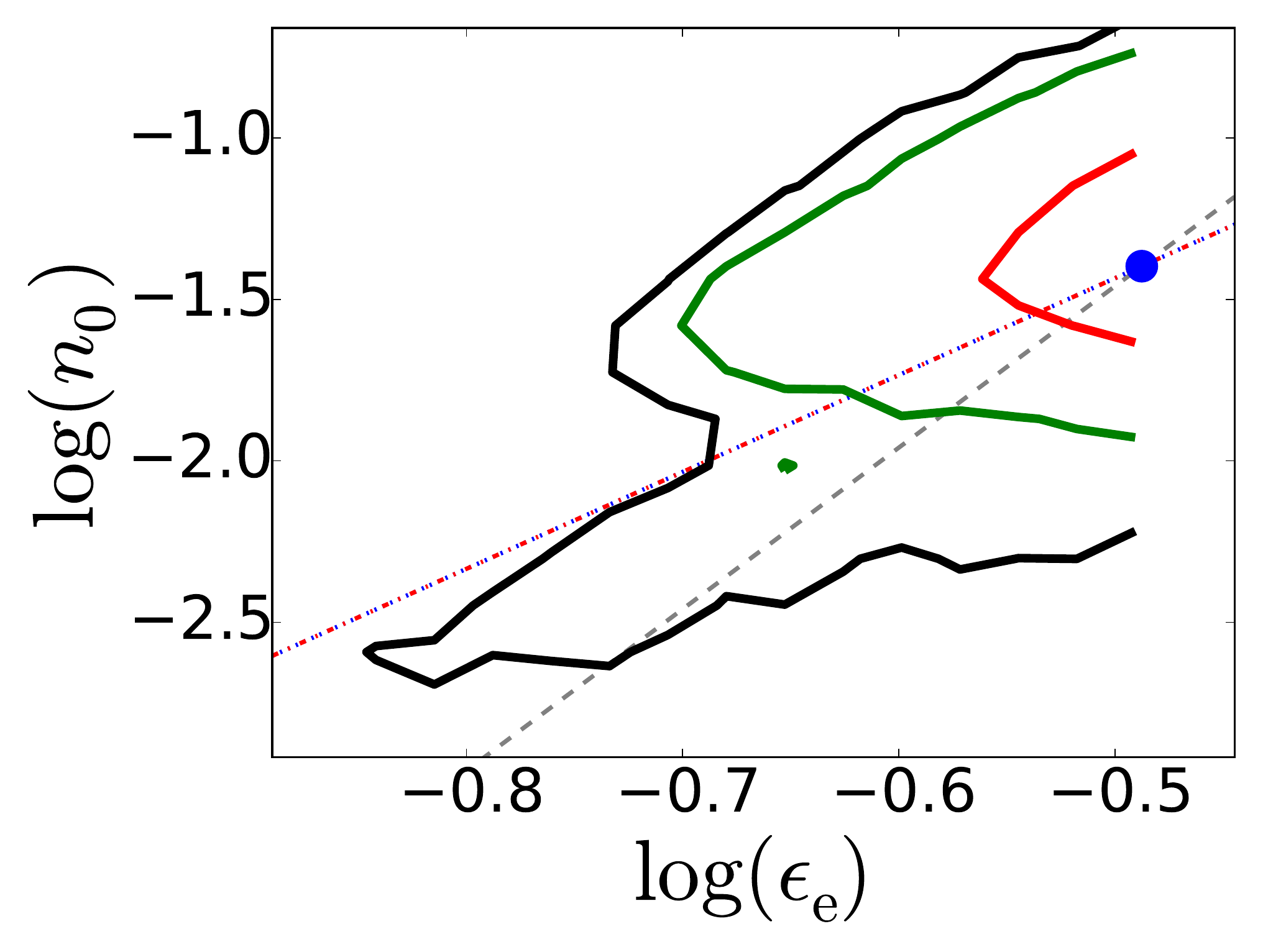} \\
 \includegraphics[width=0.30\textwidth]{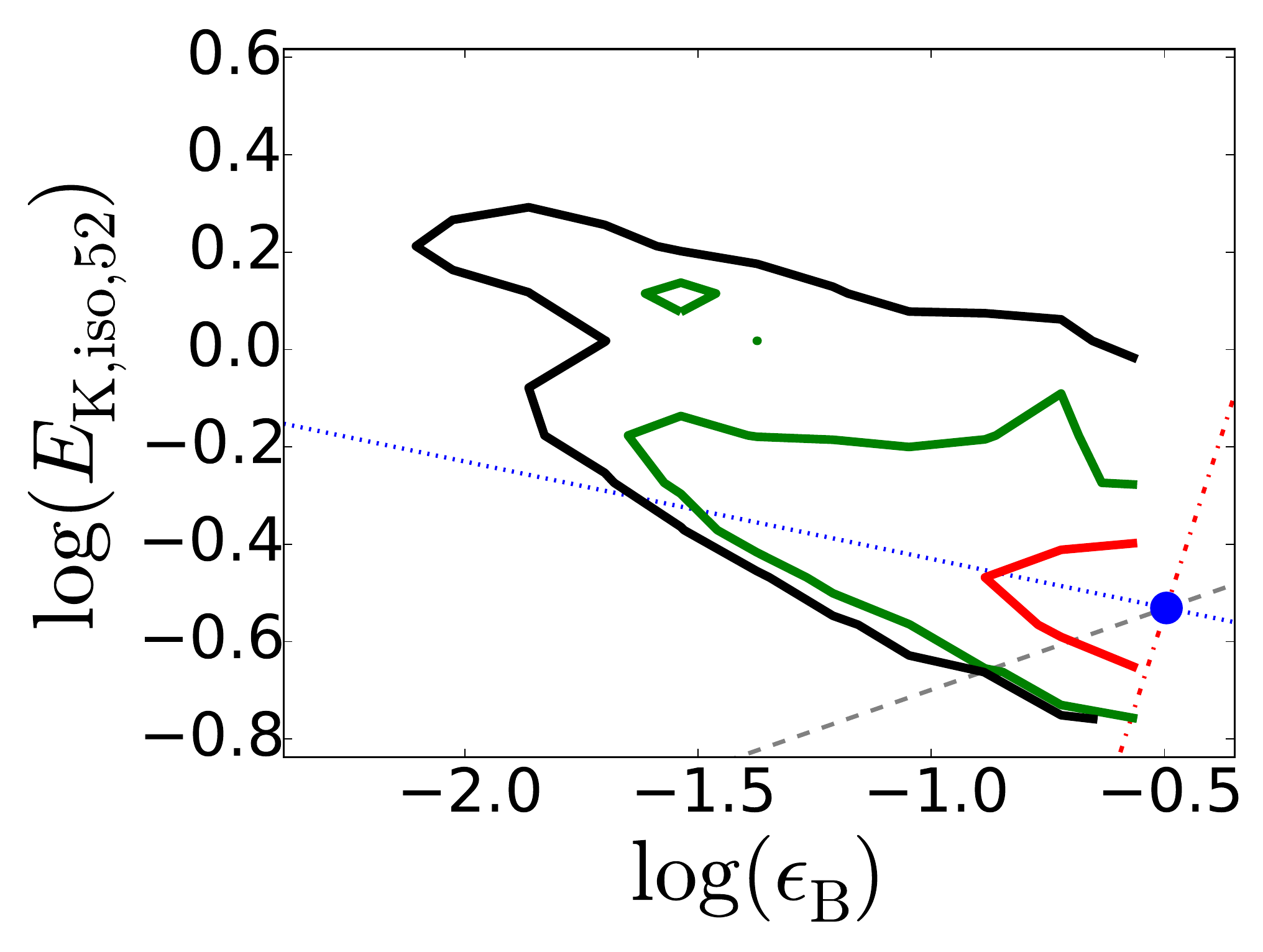} &
 \includegraphics[width=0.30\textwidth]{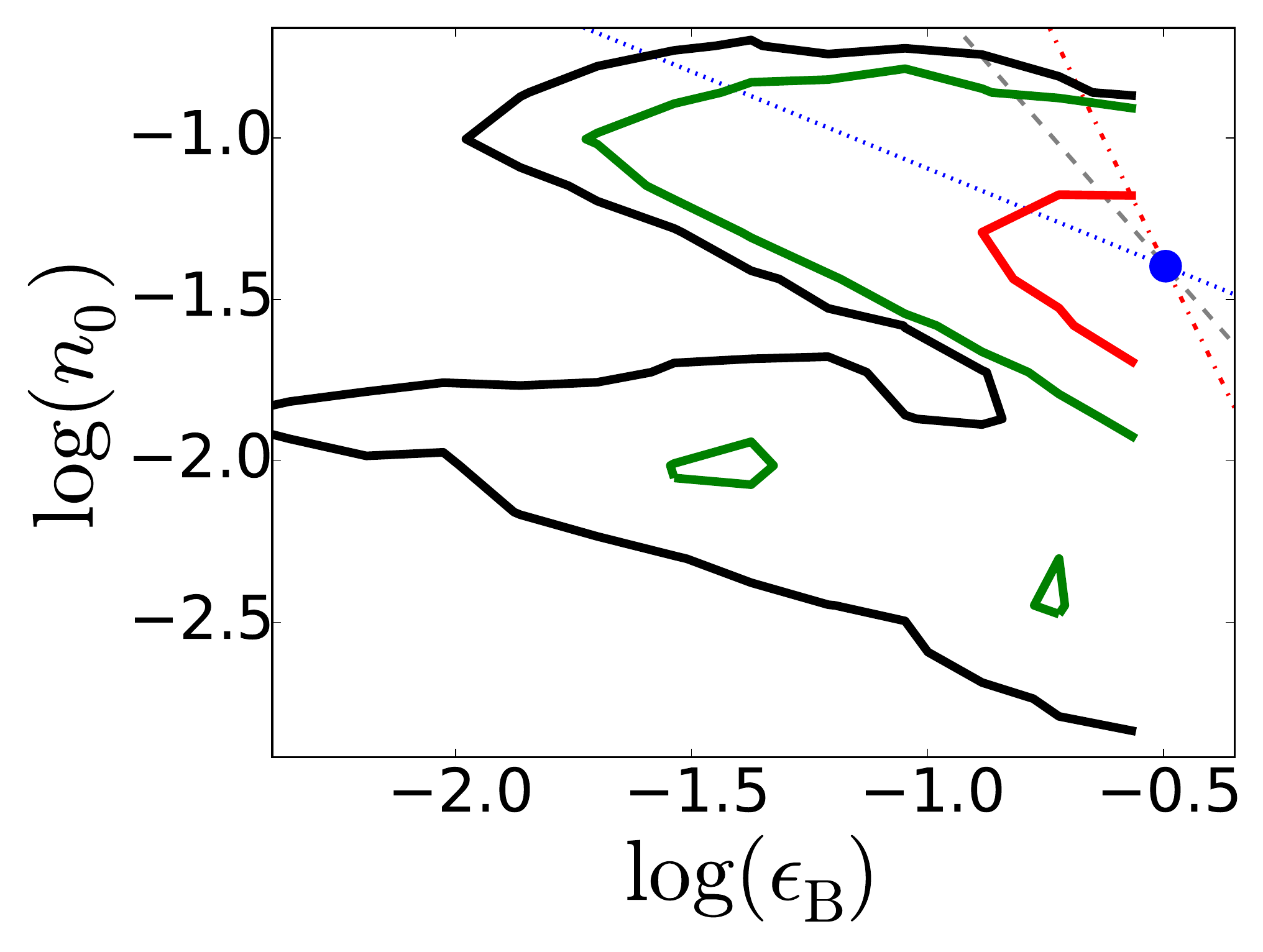} &
 \includegraphics[width=0.30\textwidth]{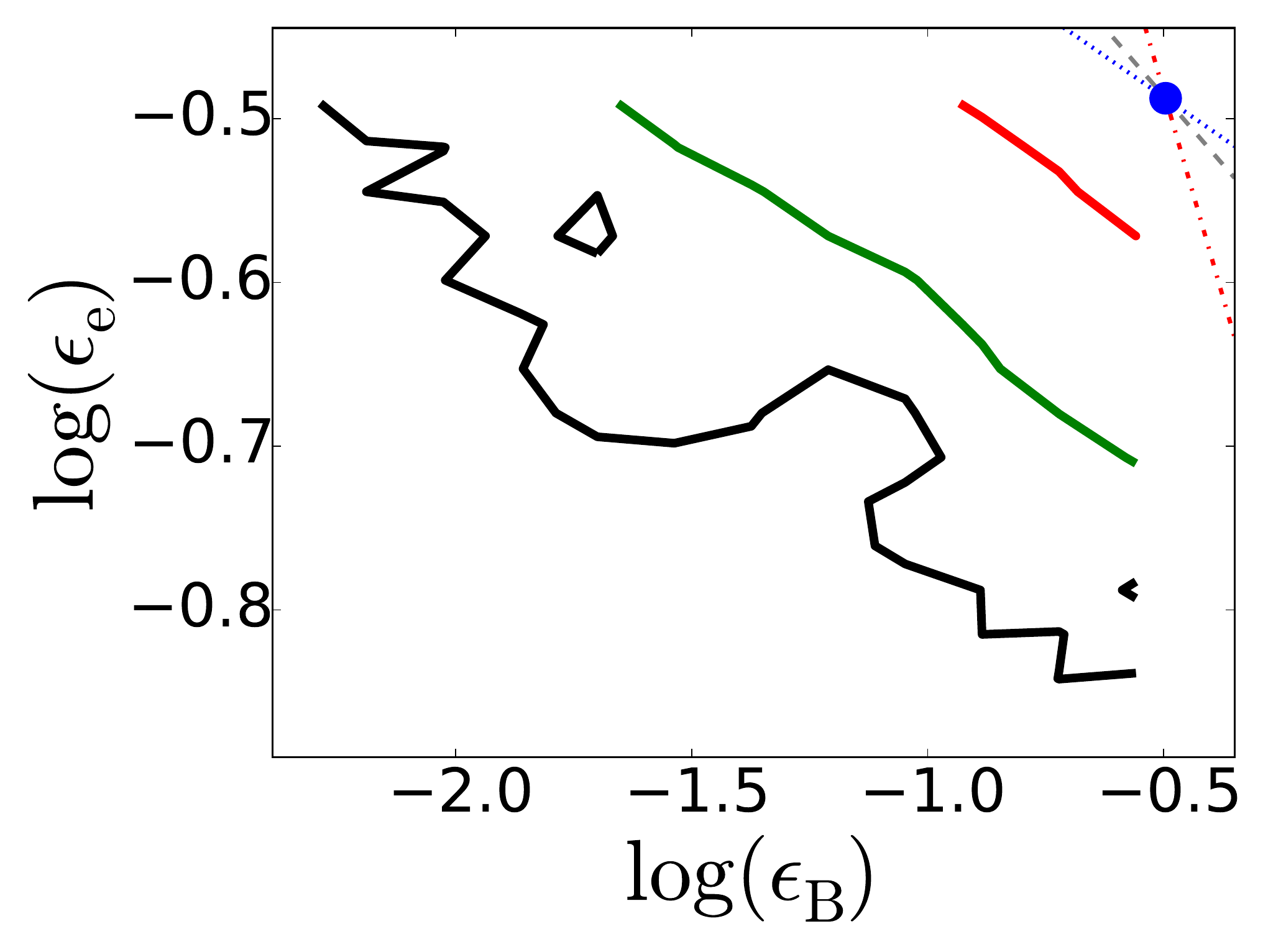} \\
\end{tabular}
\caption{1$\sigma$ (red), 2$\sigma$ (green), and 3$\sigma$ (black) contours for correlations
between the physical parameters, \EKiso, \dens, \epse, and \epsb\ for GRB\,120923A, in the ISM-like 
model from Markov chain Monte Carlo simulations. We have restricted $\epsilon_{\rm e} < 1/3$ and 
$\epsilon_{\rm B} <1/3$.  The lines indicate the expected relations between these
parameters when \nua\ (grey, dashed), \nuc\ (blue, dotted), or \numax\ (red, dash-dot) are not 
fully constrained, and are provided for reference, normalized to pass through the 
highest-likelihood point (blue dot).
} 
\label{fig:120923A_ISM_corrplots}
\end{figure*}

Applying the expression for \thetajet\ above to our MCMC chains with their 
individual values of \EKiso, \dens, $z$, and \tjet, we find $\thetajet=5.0^{+1.3}_{-0.8}$\,degrees
and $\EK=(1.2^{+0.5}_{-0.2})\times10^{49}$\,erg. Correcting $\Egammaiso$ for beaming using this 
measurement of \thetajet, we find $\Egamma=(1.8\pm0.8)\times10^{50}$\,erg. We present histograms of 
the marginalized posterior density for each parameter in Figure~\ref{fig:120923A_ISM_hists} and 
correlation contours between the physical parameters and expected relations between the 
parameters in the absence of constraints on one of the spectral break frequencies in 
Figure~\ref{fig:120923A_ISM_corrplots}. We 
summarise the results of our MCMC analysis in Table \ref{tab:param_summary}.

\begin{deluxetable}{lcc}
\tabletypesize{\footnotesize}
\tablecolumns{3}
\tablewidth{0pt}
\tablecaption{Parameters from multi-wavelength modelling of GRB\,120923A\label{tab:param_summary}}
\tablehead{
  \colhead{Parameter}   &
  \colhead{Best-fit value} &  
  \colhead{MCMC result}  \\  
  }
\startdata
$z$          & 8.1                & $8.1^{+0.2}_{-0.3}$ \\ [1ex]
$p$          & 2.5                & $2.7^{+0.3}_{-0.2}$ \\ [1ex]
\epse        & 0.33               & $0.31^{+0.02}_{-0.04}$ \\ [1ex]
\epsb        & 0.32               & $0.23^{+0.07}_{-0.11}$ \\ [1ex]
\dens  (cm$^{-3}$)      & $4.0\times10^{-2}$ & $(4.1^{+2.2}_{-1.4})\times10^{-2}$ \\ [1ex]
$E_{\rm K,iso}$ ($10^{51}$\,erg) &  $2.9$ & $3.2^{+0.8}_{-0.5}$\\ [1ex]
\tjet (d)    & 3.0                & $3.4^{+1.1}_{-0.5}$ \\ [1ex]
\thetajet (degrees) & 4.9         & $5.0^{+1.3}_{-0.8}$ \\ [1ex]
\AV (mag)    & 0.06               & $\lesssim0.08^{\dag}$\\ [1ex]
$E_{\rm \gamma, iso}^{\ddag}$ ($10^{52}$\,erg) & \multicolumn{2}{c}{$4.8^{+6.1}_{-1.6}$} \\ [2ex]
\hline
$E_{\rm \gamma}$ ($10^{50}$\,erg)  & 1.8  & $1.8\pm0.8^{*}$ \rule{0pt}{3ex}\\[1ex]
$E_{\rm K}$ ($10^{49}$\,erg)       & 1.1  & $1.2^{+0.5}_{-0.2}$\\
\enddata
\tablecomments{$^{\dag}$ 90\% confidence upper limit. The median value of the host extinction in 
our MCMC analysis is $\AV=3.8\times10^{-5}$ mag with a 68\% credible interval $\AV \in 
(0.00,0.06)$. 
$^{\ddag}$ 1--$10^4$\,keV, rest frame.
$^{*}$ Using symmetrized uncertainties (one-half positive error minus negative error) for both 
\Egammaiso\ and \thetajet, followed by a Monte-Carlo calculation.}
\end{deluxetable}

\section{Discussion}
The photometric redshift derived from SED-fitting ($z=8.1\pm0.4$) and multi-wavelength modelling 
($z=8.1^{+0.2}_{-0.3}$) agree with the redshift derived from the X-shooter spectrum 
($z=7.84^{+0.06}_{-0.12}$).  As expected, the multi-wavelength modelling produces a narrower 
posterior density function compared to SED-fitting alone, while the spectral analysis providing the 
strongest constraint of the three methods. In principle, it is possible to use the posterior density 
function of $z$ derived from the spectrum as a prior on the redshift for the multi-wavelength 
analysis. However, a perusal of the correlation contours between the redshift and the other 
parameters in the MCMC results of the multi-wavelength analysis suggests that the redshift is not 
strongly coupled to the other parameters and that, therefore, imposing such a prior is of limited 
utility. In confirmation, we find that selecting the multi-wavelength analysis MCMC samples within 
the redshift range $7.72 < z < 7.90$ (the 68\% credible interval from the spectral analysis) 
results in identical posteriors for the other parameters as for the full distribution. The lack of 
a strong correlation between $z$ and the other parameters suggests that the measurement of $z$ is 
driven by a small subset of the data, essentially in a model-independent fashion.

To place our measured value of the circumburst density for GRB\,120923A in context, we compute 
summary statistics for the circumburst density for GRBs with ISM-like environments reported and 
aggregated in \cite{Laskar14} and \cite{lbm+15}. Since the density spans several orders of magnitude 
and therefore acts as a scale parameter, we use $\varrho\equiv\log_{10}({\dens/{\rm cm^{-3}}})$ for this analysis. 
We find $\bar{\varrho} = -0.19$, the standard deviation, $\sigma_{\varrho}=2.0$, and the median, 
$\hat{\varrho}=-0.17$. In comparison, we have $\varrho=-1.4\pm0.2$ for GRB\,120923A, such that 
$|(\varrho-\bar{\varrho})|/\sigma_{\varrho}\approx0.6$. Whereas the measured density is lower than 
both the mean and median reported for GRB afterglows thus far, it is consistent with being drawn 
from the same distribution.

\begin{figure}
\includegraphics[width=0.5\textwidth]{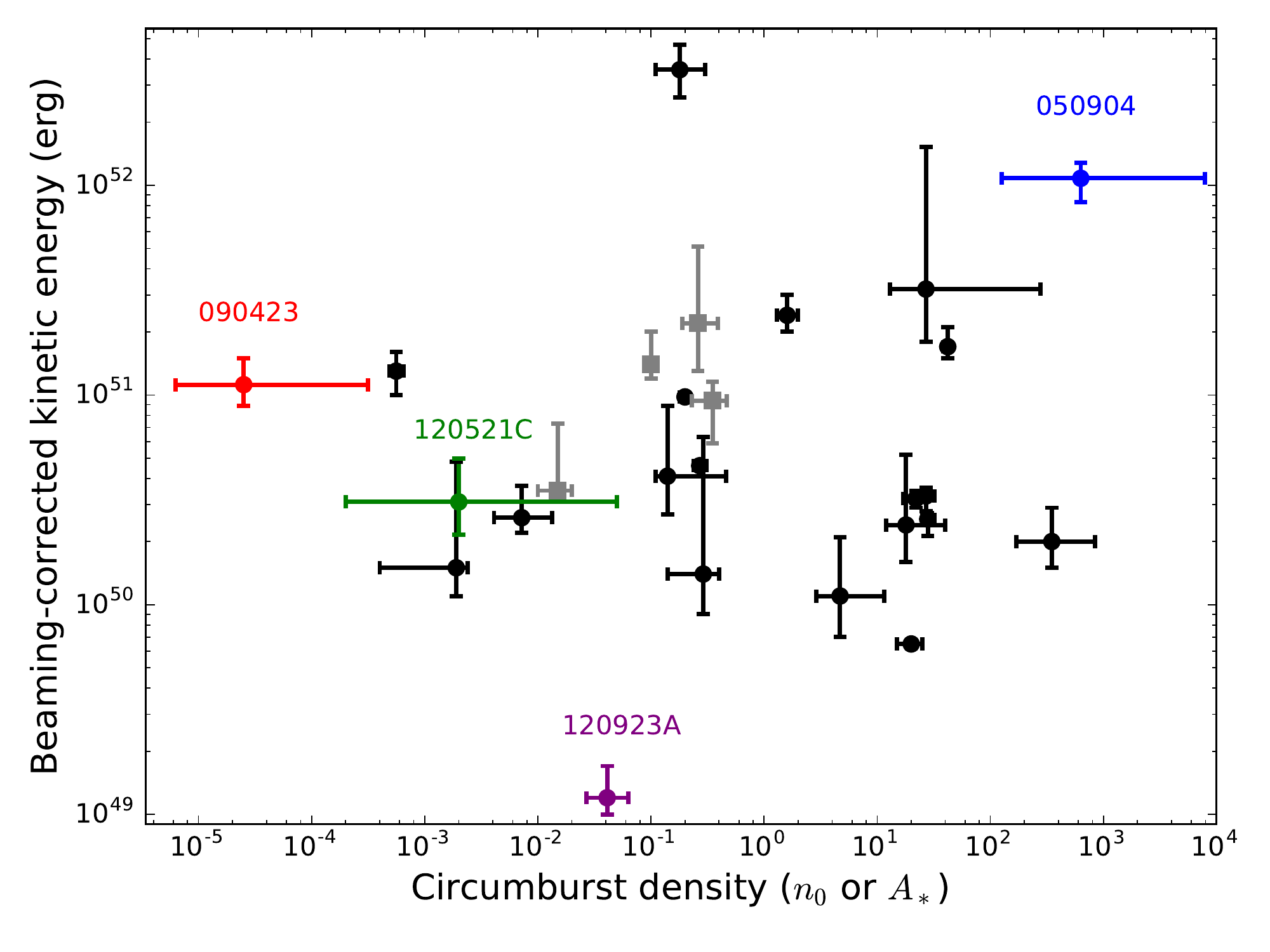}
\caption{Beaming-corrected kinetic energy as a function of circumburst density in units of cm$^{-3}$ from 
multi-wavelength modelling of GRB afterglows for both ISM (black circles) and wind-like 
environments (grey squares) at $z\sim1$ \citep[grey and black; 
][]{pk02,yhsf03,ccf+08,cfh+10,cfh+11,lbm+15} and at $z\gtrsim6$ (blue: GRB\,050904, red: 
GRB\,090423A, green: GRB\,120521C, and purple: GRB\,120923A; from \citealt{Laskar14} and this 
work).
\label{fig:En}}
\end{figure}

For the beaming corrected kinetic energy, we once again work with the logarithm:
$\varepsilon\equiv\log_{10}{(E_{\rm K}/10^{50}\,{\rm erg})}$ and have $\bar{\varepsilon}=0.75$, 
$\sigma_{\varepsilon}=0.70$, and $\hat{\varepsilon}=0.58$ for the comparison sample. For 
GRB\,120923A, we find $\varepsilon=-0.9^{+0.16}_{-0.10}$, such that 
$|(\varepsilon-\bar{\varepsilon})|/\sigma_{\varepsilon}\approx2.4$. Thus, although we cannot rule out that 
GRB\,120923A is drawn from the same sample as the comparison events based on \EK, the measured value 
of the beaming-corrected kinetic energy in the case of this event is one of the lowest observed for 
GRB afterglows so far (Figure \ref{fig:En}). A caveat here is that the comparison sample for which these
parameters have been derived consists of well-studied and generally bright events, and so could itself be 
biased compared to the wider population. 

\begin{figure}
\includegraphics[width=0.5\textwidth]{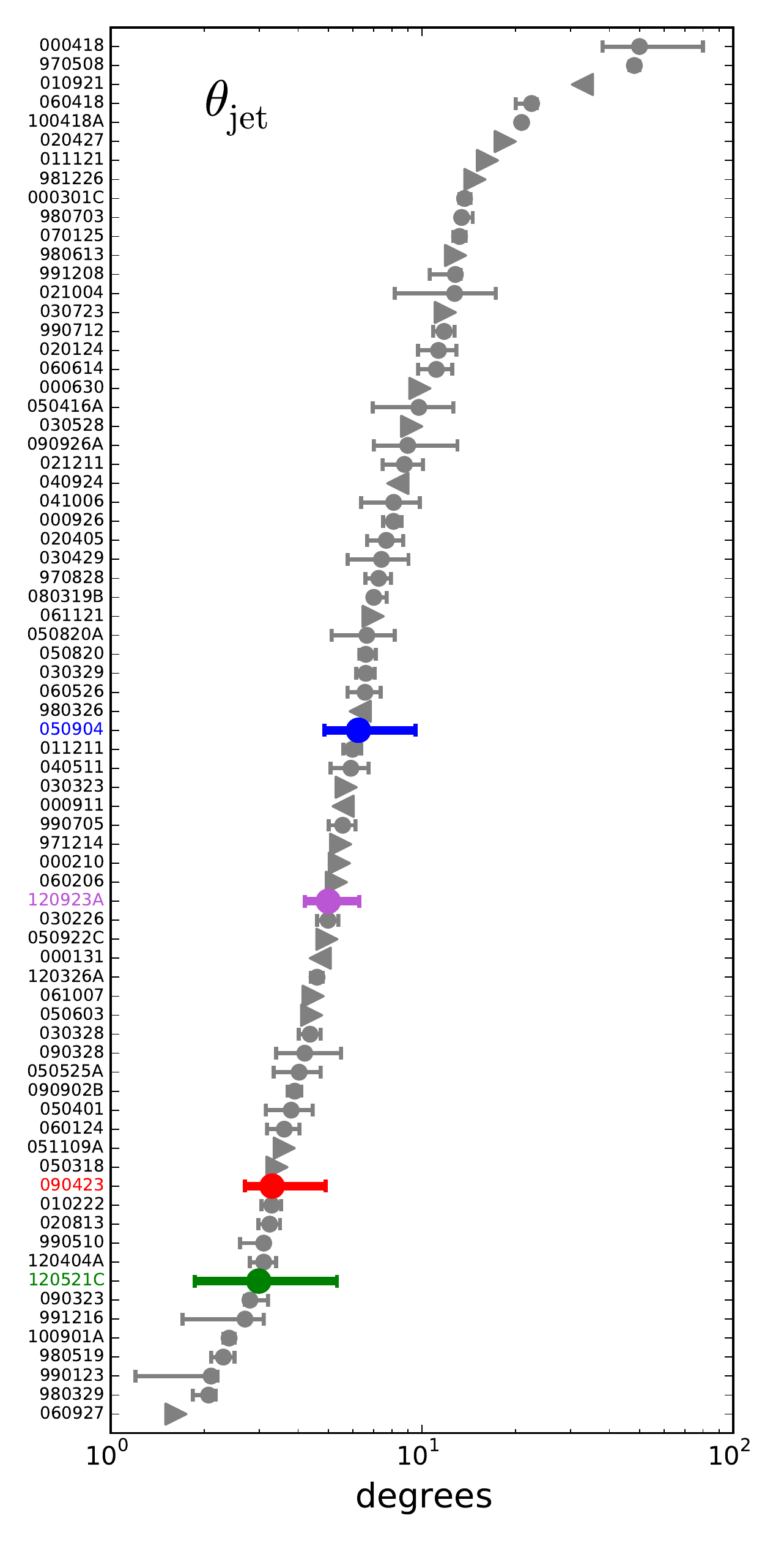}
\caption{Jet opening angles from multi-wavelength modelling of GRB afterglows at $z\sim1$ 
\citep[grey;][]{fb05,gngf07,cfh+10,cfh+11,lbm+15} and at $z\gtrsim6$ (blue: GRB\,050904, red: 
GRB\,090423A, green: GRB\,120521C, and purple: GRB\,120923A; from \citealt{Laskar14} and this 
work).
\label{fig:thetajet}}
\end{figure}

Our {\it HST} observations enabled us to measure the steep light curve decay which we have attributed to a jet break. 
Although the identification of jet breaks in GRB afterglow light curves has sometimes proven controversial in the {\em Swift} era \citep[e.g.,][]{curran08},
the sharp and marked break in this case is rather hard to interpret otherwise.
Through multi-wavelength modelling, we have derived a jet opening angle of 
$\thetajet=5.0^{+1.3}_{-0.8}$\,degrees, the fourth such measurement at $z\gtrsim6$. Interestingly, this 
value is comparable to the values obtained for other $z\gtrsim6$ events, 
but is smaller than the median value for $z\sim1$ events reported in the literature (Figure~\ref{fig:thetajet}; 
although we caution that limited data for many older bursts in this compilation means that interpretation of
temporal breaks as being due to beaming is less secure). 
This supports the hypothesis of \cite{Laskar14} that observed
high-redshift GRBs may be more tightly-beamed on the average than their more local counterparts, which may
be a consequence of narrower jets leading to more intrinsically luminous and hence easier to observe afterglows.
Multi-wavelength analysis for more high-redshift events coupled with a uniform statistical study of 
the $z\sim1$ events would further clarify this inference.

\section{Conclusions}
We have presented X-ray, NIR, and radio observations of GRB\,120923A. The faintness of the
afterglow made the initial identification as an optical drop-out and subsequent spectroscopy
challenging. Nonetheless, we were able to derive a redshift for the event of 
$z=7.84^{+0.06}_{-0.12}$ from a low signal-to-noise VLT/X-shooter spectrum, which is consistent with
that obtained from the photometric redshift analysis. 
The absence of significant flux
at the afterglow location in our final {\em HST} image suggests the host galaxy 
is likely fainter than $M_{\rm F140W,AB}\gtrsim27.5$, consistent with the deep limits on other
$z\sim8$ GRB hosts \citep{Tanvir12}.

Our multi-wavelength modelling of all available afterglow observations, shows that a standard external shock
in a constant-density circumburst 
environment with $\dens\approx0.04\,{\rm cm^{-3}}$ explains the data well. Using deep {\em HST} observations, we find 
evidence for a jet break at $\tjet=3.4^{+1.1}_{-0.5}$\,days, from which we computed a jet opening angle of 
$\thetajet=5.0^{+1.3}_{-0.8}$\,degrees. Our results support the apparent trend of smaller opening angles for 
$z\gtrsim6$ GRBs compared to $z\sim1$ events.
This may reflect the fact that  at high redshift we can only  detect events with the highest isotropic luminosities,
which would therefore favour selection of more narrowly beamed jets assuming a fixed range of intrinsic energy reservoirs.
The blastwave kinetic energy, 
$E_{\rm K}=1.2^{+0.5}_{-0.2}\times10^{49}$\,erg, is one of the lowest seen so far for both nearby and high-$z$ well-studied events.
Otherwise the properties of GRB\,120923A, like those of the other $z\gtrsim6$ bursts discovered to date \citep{Laskar14}, show no signatures
that would suggest they could be produced by Pop III stars, such as very long duration or extremely large
energy \citep[cf.][]{Meszaros2010}.

In the case of GRB\,120923A, NIR observations within the first hours post-burst alerted us to the high-redshift 
nature of this event. In addition, they were essential to catch the peak of the afterglow SED at a 
time that the radiation was in the fast cooling regime, allowing us to constrain the circumburst 
density even in the absence of a radio detection and the resulting freedom in locating the 
synchrotron self-absorption frequency. Rapid-response NIR observations at large telescopes are 
therefore crucial not only for their ability to help us identify GRBs at $z\gtrsim6$, but 
also for studying the progenitors and environments of these energetic phenomena, establishing them 
as unique probes of star-formation at the highest redshifts.  In the {\em JWST} era, NIR spectroscopy,
even several days post-burst, of similar events will provide much higher signal-to-noise data,
allowing meaningful constraints to be placed on abundances and neutral hydrogen in the host galaxy.

\acknowledgments

We thank the anonymous referee for their constructive comments.

This work is based on observations made with the NASA/ESA Hubble Space Telescope, obtained at the 
Space Telescope Science Institute, which is operated by the Association of Universities for Research 
in Astronomy, Inc., under NASA contract NAS 5-26555. These observations are associated with program 
GO12558. Support for Program number GO12558 was provided by NASA through a grant from the Space 
Telescope Science Institute, which is operated by the Association of Universities for Research in 
Astronomy, Incorporated, under NASA contract NAS5-26555.

This work is based on observations collected at the European Organisation for Astronomical 
Research in the Southern Hemisphere (ESO), Chile under programme 089.A-0067, and on  
observations obtained at the Gemini Observatory (acquired through the Gemini Science Archive and 
processed using the Gemini IRAF package), which is operated by the  Association of Universities for 
Research in Astronomy, Inc., under a cooperative agreement with the NSF on behalf of the Gemini 
partnership: the National Science Foundation (United States), the National Research Council 
(Canada), CONICYT (Chile), the Australian Research Council (Australia), Minist\'{e}rio da 
Ci\^{e}ncia, Tecnologia e Inova\c{c}\~{a}o (Brazil) and Ministerio de Ciencia, Tecnolog\'{i}a e 
Innovaci\'{o}n Productiva (Argentina). The National Radio Astronomy Observatory is a facility of the 
National Science Foundation operated under cooperative agreement by Associated Universities, Inc.

When the data  reported here were acquired, UKIRT was operated by the Joint Astronomy Centre on behalf of the Science and Technology Facilities Council of the U.K. We thank Tim Carroll for his assistance in making these observations.

Based on data obtained with the VLA under program 12A-394. 
The National Radio Astronomy Observatory is a facility of the National Science Foundation operated under cooperative agreement by Associated Universities, Inc.

The Dark Cosmology Centre was funded by the DNRF. The research leading to these results has received 
funding from the European Research Council under the European Union's Seventh Framework Program 
(FP7/2007-2013)/ERC Grant agreement no.  EGGS-278202.

DAK is grateful to TLS for financial support.

ERS acknowledges support from UK STFC consolidated grant  ST/L000733/1

DM thanks the Instrument Center for Danish Astrophysics (IDA) for support.

AdUP acknowledges support from a Ram\'on y Cajal fellowship.

RSR and AdUP acknowledge support from a 2016 BBVA Foundation Grant for Researchers and Cultural Creators. 

DAK, ZC, RSR and AdUP acknowledge support from the Spanish research project AYA 2014-58381-P.

DX acknowledges the support by the One-Hundred-Talent Program of the
Chinese Academy of Sciences (CAS), by the Strategic Priority Research
Program ÒMulti-wavelength Gravitational Wave UniverseÓ of the CAS (No.
XDB23000000), and by the National Natural Science Foundation of China
under grant 11533003.

TK acknowledges support through the Sofja Kovalevskaja Award to Patricia Schady from the Alexander von
Humboldt Foundation of Germany.

NRT and KW acknowledge support from the UK STFC under consolidated grant 
ST/N000757/1.

%

\vspace{5mm}
\facility{Gemini-North (GMOS, NIRI)}, 
\facility{HST (WFC3-IR)}, \facility{VLT (X-shooter, ISAAC, FORS2)}, \facility{UKIRT (WFCAM)}, 
\facility{VLA}

\end{document}